\documentclass[a4paper,12pt]{article}

\usepackage{palatino}
\usepackage{soul}
\usepackage{color}
\usepackage{multirow}
\usepackage{amsmath,amssymb}
\usepackage{amsthm}
\usepackage{dirtytalk}
\usepackage{graphicx}
\usepackage{enumitem}
\usepackage{setspace}
\usepackage{booktabs}
\usepackage{epstopdf}
\usepackage{footnote}
\usepackage{geometry}
\usepackage{rotating}
\usepackage{titlesec}
\usepackage{amsmath,bm}
\usepackage{lipsum} 
\usepackage{tabularx}
\usepackage{mathtools}
\usepackage{subfigure}
\usepackage{textcomp}
\usepackage{geometry}
\usepackage{amsmath}
\usepackage{amsfonts}
\usepackage{amssymb}
\usepackage{mathtools}
\usepackage{graphicx}
\usepackage{verbatim}
\usepackage[round,authoryear]{natbib}
\usepackage{tabularx}
\usepackage{graphicx} 
\usepackage{caption,setspace}
\usepackage[capposition=top]{floatrow}
\usepackage{ragged2e}
\usepackage{multirow}
\usepackage{hhline}
\usepackage{colortbl}
\usepackage{multicol}
\usepackage{xcolor}
\usepackage{breqn}
\usepackage{booktabs}
\usepackage[toc]{appendix}
\usepackage[colorlinks=true,linkcolor=blue,citecolor=black,urlcolor=red]{hyperref}
\usepackage{bookmark}
\usepackage{placeins}
\usepackage{bm}
\usepackage[usestackEOL]{stackengine}
\usepackage[normalem]{ulem}
\usepackage{titling}

\newcolumntype{P}[1]{>{\centering\arraybackslash}p{#1}}
\newcolumntype{L}[1]{>{\arraybackslash}p{#1}}
\makeatletter
\def\@biblabel#1{\hspace*{-\labelsep}}
\makeatother
\graphicspath{{Figures/}}
\usepackage{palatino}
\usepackage[disable]{todonotes}
\title{\vspace{-2.5cm} The Unequal Costs of Pollution: \\ Carbon Tax, Inequality, and Redistribution\thanks{We wish to thank participants to the Sapienza Macro Workshop for their comments and Salvatore Nisticò for the discussion.}}
\author{Cristiano Cantore \thanks{Sapienza University of Rome. Email: \url{cristiano.cantore@uniroma1.it}.}
\and
Giovanni Di Bartolomeo \thanks{Sapienza University of Rome. Email: \url{giovanni.dibartolomeo@uniroma1.it}.}
\and
Francesco Saverio Gaudio \thanks{Sapienza University of Rome. Email: \url{francescosaverio.gaudio@uniroma1.it}. \vspace{0.25cm}}}

\date{February 26, 2025} 
\begin{document}
\maketitle
\thispagestyle{empty}

\begin{abstract}
\noindent This paper studies how household heterogeneity affects the level and cyclical behavior of the optimal carbon tax in a real economy. We demonstrate that an equity-efficiency trade-off arises due to income inequality and heterogeneity in the marginal disutility of pollution. Two scenarios are analyzed: one with unrestricted income redistribution to mitigate inequality and another where redistribution is constrained to carbon tax revenues. Our findings reveal that household heterogeneity and redistribution policies significantly shape the level and cyclical behavior of the optimal carbon tax, decoupling it from the social cost of carbon. When the planner prioritizes redistribution towards poorer households, the optimal tax rate is lower than in the unconstrained scenario, and its fluctuations are amplified by countercyclical inequality. 
\vspace{0.1cm}\\ 
\end{abstract}

 \newpage

\pagenumbering{arabic} 


\section{Introduction}
\label{sec:Intro}
Climate change poses one of the most pressing challenges of our time, with carbon taxes widely recognized as a cost-effective tool for reducing greenhouse gas emissions. However, designing optimal carbon taxes is a complex task that must account for the heterogeneous effects of these policies across diverse households. Differences in income, wealth, and access to financial markets significantly influence carbon taxation's economic and social impacts, shaping its feasibility and effectiveness. For instance, low-income households, which allocate a larger share of their income to energy-intensive goods, are disproportionately vulnerable to rising energy costs induced by carbon taxes. These regressive effects risk exacerbating existing income inequalities unless carefully addressed.\footnote{Among others, \cite{grainger2010}, \cite{Sager19}, and \cite{belfiori2024unequal} highlight how the emissions associated with household spending differ across income levels. See also \cite{Bento2009} and \cite{FullertonHeutel10}, who focus on carbon taxes.} 

Motivated by the need to balance environmental objectives with social equity, this paper investigates the optimal design of carbon taxes within a dynamic stochastic general equilibrium (DSGE) framework. The analysis focuses on the interactions between carbon taxation, household heterogeneity, and macroeconomic dynamics. Specifically, we address several key questions: How does household heterogeneity shape the level and structure of optimal carbon taxes? What are the trade-offs between equity and efficiency in economies characterized by significant income and consumption inequality? And what are the implications of household heterogeneity for the cyclical behavior of the social cost of carbon and optimal taxation?

Our approach is built on four main blocks. First, we consider a real economy with limited asset market participation where carbon emissions represent a negative externality in the households' utility function that firms do not internalize, with the optimal response being a Pigouvian tax. Limited asset market participation gives rise to consumption and income heterogeneity between "hand-to-mouth" (henceforth HtM) households---who consume labor income and transfers---and "savers"---who have full access to financial markets. As a result, only a fraction of households own the productive sector, implying that taxes on firms' production are effectively levied on the financial income accruing to savers \citep[similar to][]{kanzig2023unequal}.

Second, a central assumption is that consumer utility is non-separable in (higher) consumption and (lower) emissions, implying that the negative effects of emissions depend on the household-specific level of consumption.\footnote{Formally, the model incorporates the environmental externality in a non-additive manner, implying that consumption and the stock of emissions are complementary \citep[referred to as the compensation effect by][]{michel1995disutility}.} This generates inequality in the marginal disutility of pollution and allows us to formalize the stylized fact that poorer households suffer relatively more from environmental hazards \citep{eea2018unequal,hausman2021inequality, cain2024environmental,colmer2024income}.\footnote{An alternative setup would include the climate externality as directly affecting production rather than utility. However, our modeling choice seems more natural, given our focus on heterogeneity in the effects of pollution across households. Furthermore, including the externality in utility better fits the notion of conventional pollutants that directly affect health \citep[as also discussed in][]{heutel2012how}.} In this respect, our notion of HtM is intended to capture poorer households \citep[rather than the "wealthy" HtM, as in][]{kaplan2014wealthy}, consistent with the aforementioned studies on the differential exposure of the bottom of the income distribution to the adverse effects of climate change.\footnote{Relatedly, we verify the sensitivity of our quantitative results to different calibrations of the parameter controlling the HtM population share.}

Third, we analyze two policy scenarios for designing the optimal carbon tax. In the unconstrained scenario, the planner can redistribute income freely across households, eliminating consumption and income inequality. Here, the optimal tax aligns with the social cost of carbon (SCC), resembling a representative agent framework and achieving the first-best outcome. In the constrained scenario, redistribution is limited to tax revenues, leading to consumption and income inequality in equilibrium. In this setting, the planner faces an efficiency-equity trade-off, as the tax must balance the dual objectives of addressing environmental externalities and mitigating inequality. This scenario introduces key differences: the SCC incorporates the elevated emission disutility experienced by poorer households, while the planner's redistributive motive affects the private cost of emissions and the resulting optimal tax rate.

Finally, we examine how economic fluctuations generated by total factor productivity (TFP) shocks influence optimal carbon tax rates and their interaction with household welfare. Specifically, we explore how countercyclical consumption inequality mitigates fluctuations in the SCC by disproportionately benefiting poorer households during economic expansions. Redistribution policies further shape the cyclical behavior of the tax, with fluctuations in inequality introducing a time-varying wedge between the SCC and the tax. 

Our findings contribute to the literature on climate policy by emphasizing the role of household heterogeneity and macroeconomic fluctuations in shaping the optimal design of carbon taxes. First, we show that household heterogeneity significantly influences the optimal carbon tax, both in its level and cyclical behavior. The presence of HtM households, who consume their entire income and display a higher marginal disutility from the environmental externality, increases the social cost of carbon and leads to a higher tax rate than a representative agent framework. Second, redistributive constraints play a critical role in shaping the optimal tax rate. When redistribution is limited to tax revenues, the tax rate reflects both the environmental damage from emissions and the potential to alleviate inequality. Third, we find that the cyclical properties of inequality affect the dynamics of the carbon tax. Countercyclical inequality dampens SCC fluctuations, but redistribution motives amplify the tax's cyclical variability. This is most evident when the planner prioritizes redistribution towards poorer households, which, while reducing inequality, introduces greater volatility in the tax rate.

\paragraph{Related literature} 
Our paper contributes to the growing body of research \citep[see, among others,][]{heutel2012how,golosov2014optimal,annichiarico2015policy,benmir2020green,gibson2023pollution,kanzig2023unequal,sahuc2024climate} addressing climate issues through business-cycle models.\footnote{For a review of the literature on business cycles and climate policies, see \citet{annichiarico2022primer}.} While most of these studies adopt a representative-agent framework \citep[except for][]{kanzig2023unequal}, we advance the literature by introducing a limited asset market participation model that underscores the quantitatively significant role of cyclical inequality in shaping the optimal carbon tax. Specifically, we demonstrate that the optimal tax is not necessarily procyclical \citep[as suggested, for instance, by][]{heutel2012how,benmir2020green} when consumption and income inequality exhibit sufficient countercyclicality, particularly under a scheme where tax revenues are rebated to firm owners. Moreover, our results reveal that the optimal average tax in the two-agent model diverges significantly from its representative-agent counterpart, varying based on the redistribution scheme employed.

This work closely aligns with the contribution of \citet{kanzig2023unequal}, who examines the heterogeneous effects of carbon pricing in a two-agent framework similar to ours. The author argues that redistributing carbon tax revenues to the poorer---who bear the most significant burden due to their higher share of spending on energy-intensive goods—--can mitigate its adverse economic impacts by stabilizing inequality fluctuations while still achieving substantial emission reductions. In a similar vein, we demonstrate that the constrained optimal policy with redistribution to the HtM reduces and stabilizes inequality (in terms of absolute deviations from the steady state) relative to the business-as-usual (BAU) scenario while prompting firms to cut emissions by $20\%$ to $30\%$. However, unlike \citet{kanzig2023unequal}, who evaluates various redistributive schemes under a given (and not necessarily optimal) tax rate, our approach reveals a critical trade-off between environmental and redistributive objectives in the design of the optimal carbon tax.

Despite abstracting from nominal rigidities and focusing on the issue of optimal carbon taxation \citep[similar to][in a representative-agent setup]{heutel2012how,benmir2020green}, our work also connects to a broader macroeconomic literature exploring the interrelations between limited asset market participation, inequality fluctuations, and business cycles \citep[e.g.,][]{bilbiie2020new,bilbiie2024monetary,cantore2021workers,bilbiie2022inequality}. Specifically, we highlight how consumption inequality between savers and hand-to-mouth households introduces a wedge in the social planner's Euler equation for the stock of greenhouse gases. This wedge, whose level and dynamics strictly depend on alternative redistribution schemes for carbon tax revenues, influences both the first and second moments of the social cost of carbon and, consequently, the optimal carbon tax.

Our finding that inequality and redistribution have an impact not only on the \textit{level} but also the \textit{dynamics} of the optimal carbon tax in response to aggregate shocks constitutes a key contribution to the growing literature on carbon taxation in heterogeneous agent settings \citep{fullerton2013pollution,jacobs2019redistribution,goulder2019impacts,kanzig2023unequal,fried2024understanding}. Indeed, most studies in this field primarily focus on the tax level and abstract from aggregate uncertainty in finding the transition path from a BAU to a (potentially optimal) carbon tax scenario. Our analysis differs in other aspects as well. First, unlike \citet{fried2024understanding} and \citet{belfiori2024unequal}, who use Stone-Geary preferences to capture the stylized fact that poorer households allocate a higher share of their expenditure to energy-intensive goods, we adopt an external habit utility specification \citep[as in][]{benmir2020green} that reflects the regularity that poorer households are disproportionately affected by environmental risks, in line with the environmental justice literature reviewed by \citet{cain2024environmental}. As a result, the optimal tax in our unequal economy scenario—when rebates are provided to savers—is considerably higher than the representative-agent counterpart, which contrasts with the findings of \citet{belfiori2024unequal}.\footnote{This result aligns with \citet{jacobs2019redistribution}, who demonstrate that the Pigouvian tax rate is larger when pollution disproportionately harms the poor.} Additionally, in our model, the emission tax is levied on firm production rather than household consumption, meaning that the fiscal burden falls primarily on wealthier households, who own the entire productive sector due to limited participation in asset markets \citep[as in][]{kanzig2023unequal}. In this sense, the uniformity and progressivity of the tax rate arise by construction in our setup. Conversely, they represent a constraint and a result (respectively) in the social planner's problem in \citet{belfiori2024unequal}.

\paragraph{Structure} The rest of the paper is organized as follows. Section \ref{sec:Model} presents the limited participation setup. Sections \ref{sec:Dec Economy} and \ref{sec:Opt Tax} discuss the main analytical results, which characterize the decentralized economy and the optimal carbon tax under different redistribution schemes. In Section \ref{sec:Quant Results}, the model is calibrated and simulated to evaluate the quantitative relevance of inequality for the optimal tax, and the robustness of the main results is verified in Section \ref{sec:Robustness}. Finally, Section \ref{sec:Conclusions} concludes.

\section{Model} \label{sec:Model}

We consider an RBC model with an environmental externality due to an accumulated pollution stock, similar to \cite{heutel2012how}. Our economy, however, features two types of households: HtM, who consume all their income, and savers, who invest in financial markets. Furthermore, given our focus on the household side of the economy, the environmental externality is introduced in the utility---rather than the production---function \citep[as in][]{acemoglu2012environment,barrage2020optimal,benmir2020green,belfiori2024unequal}. Firms produce output, generating greenhouse gas (GHG) emissions. We consider both a decentralized and a centralized economy in scenarios with and without redistribution, determining the optimal carbon tax based on household inequality and its impact on the economy. In the centralized case, the government imposes a carbon tax to internalize the SCC and may redistribute resources between households.

\subsection{Stock of GHGs and emissions}

The accumulation of greenhouse gases (GHGs) depends on human activity according to the following law of motion:
\begin{equation}
    X_{t+1}=\eta X_{t}+E_{t}, \label{eq:GHGLoM}
\end{equation}
where $X_{t}$ is the concentration of GHGs in the atmosphere at time $t$, $E_{t}\geq 0$ is the flow of emissions (specified below), and $\eta\in(0,1)$ is the linear rate of continuation of CO2-equivalent emissions on a quarterly basis. 

Emissions result from human activity as follows:
\begin{eqnarray}
    E_{t} &=&(1-\mu_{t})  \phi_{1}Y_{t}^{1-\phi_{2}} , \label{eq:Em}
\end{eqnarray}
where $\mu_{t}\in[0,1]$ is the fraction of emissions abated by firms, $Y_{t}$ denotes aggregate output, and $\phi_{1}$, $\phi_{2}\geq 0$ are carbon intensity parameters governing the relationship between the production process and emissions.

\subsection{Firm} The production side of the economy is standard. Firms maximize profits in a competitive market and produce according to a Cobb-Douglas production function that combines capital and labor:
\begin{eqnarray}
    Y_{t} &=& \varepsilon_{t}^{A}AK_{t}^{\alpha}N_{t}^{1-\alpha}, \label{eq:ProdF}
    \\
    \log(\varepsilon_{t}^{A}) &=& \rho_{A}\log(\varepsilon_{t-1}^{A})+\eta_{t}^{A},
\end{eqnarray}
where $\alpha\in(0,1)$ denotes the capital share of income, $A$ is a (constant) productivity shifter, and $\varepsilon_{t}^{A}$ is a TFP shock that follows an AR(1) process in logs, with $\eta_{t}^{A}\sim N(0,\sigma_{\eta^{A}})$.

Dividends are defined as:

\begin{equation}
    D_{t} = Y_{t}-W_{t}N_{t}-I_{t}-f(\mu_{t})Y_{t}- \tau_{t}E_{t},
\end{equation}
that is output net of the wage bill, investment, abatement costs, and the (potential) tax on emissions levied by the fiscal authority. The abatement cost function $f(\mu_{t})=\theta_{1} \mu_{t}^{\theta_{2}}$ depends on $\theta_{1}$---which pins down the steady-state fraction of abated emissions, the level of abatement $\mu_{t}$ and $\theta_{2}$---the elasticity of abatement costs to abatement effort.

The capital stock follows the law of motion:
\begin{equation}
    K_{t+1} =(1-\delta) K_{t}+\Phi \left(\frac{I_{t}}{K_{t}}\right) K_{t}, \label{eq:KLoM}
\end{equation}
with $\Phi \left(\frac{I_{t}}{K_{t}}\right)$ being the capital adjustment cost function defined as in \citet{jermann1998asset}:
\begin{eqnarray}
    \Phi \left(\frac{I_{t}}{K_{t}}\right) &=& \left[\frac{b_{1}}{1-\epsilon}\left(\frac{I_{t}}{K_{t}}\right)^{1- \epsilon }+b_{2}\right]. \label{eq:KAdjC}
\end{eqnarray}
The elasticity $\epsilon$ controls the strength of adjustment costs, while the parameters $b_{1}$, $b_{2}$ ensure that adjustment costs do not affect the economy's steady state.

\subsection{Households}

We distinguish between two groups of households. A fraction of the population consists of HtM households (denoted by the superscript "H") who are assumed to be excluded from financial markets and consume their labor income every period. The rest of the population (representing a fraction $1-\gamma$) consists of savers (denoted by the superscript "S") who work, trade bonds, and own the productive sector through stock shares. To exclusively focus on heterogeneity in consumption and the disutility from pollution, we assume that both groups do not value leisure and supply their whole-time endowment to work in the firm.

\paragraph{Utility function} Both groups of households feature external habit preferences in the spirit of \citet{campbell1999force}. Specifically, both groups of households display power utility derived from consumption net of the environmental externality:
\begin{equation}
    u_{t}^{i} = \frac{(C_{t}^{i}-\chi X_{t})^{1-\sigma}}{1-\sigma}, \quad i\in\{H,R\},
\end{equation}
where $\chi\in[0,1]$ denotes the sensitivity of utility to the GHGs stock, or equivalently the strength of the environmental externality.

This specification introduces the environmental externality in a non-additive way, implying that consumption and the stock of emissions are complements \citep[the so-called \emph{compensation effect} in][]{michel1995disutility}. It accounts for the stylized fact that poorer households suffer relatively more from environmental hazards \citep{eea2018unequal,hausman2021inequality, cain2024environmental,colmer2024income}. To see this, note that the marginal disutility of an increase in the stock of GHGs is given by:
\begin{equation}
    u_{X,t}^{i}=-\chi(C_{t}^{i}-\chi X_{t})^{-\sigma}. \label{eq:MargDis_X}
\end{equation}
Hence, consumption heterogeneity implies unequal marginal disutility of pollution, even if the deep parameters are homogeneous across households. Specifically, as HtM households consume, on average, less than savers, the former will display a larger marginal disutility of accumulated GHGs.

\paragraph{HtM Households} HtM households are excluded from financial markets and only choose consumption to maximize lifetime utility:
\begin{equation}
    \underset{\{ C_{t}^{H}\} }{\max} \quad U_{t}^{H} \equiv \mathbb{E}_{0}\sum_{t=0}^{\infty}\beta^{t}\frac{(C_{t}^{H}-\chi X_{t})^{1-\sigma }}{1-\sigma }, \label{eq:HtMBC}
\end{equation}
subject to the budget constraint:
\begin{equation}
    W_{t}N_{t}^{H}+T_{t}^{H} = C_{t}^{H}.
\end{equation}
Therefore, labor income and lump-sum transfer from the government finance consumption expenditures. If negative, $T_{t}^{H}$ denotes a lump-sum tax paid to the government.

\paragraph{Savers} Savers have full access to financial markets and choose consumption, bond holdings, and stock holdings to maximize lifetime utility:
\begin{equation}
    \underset{\{ C_{t}^{S},B_{t+1},S_{t+1}\} }{\max } \quad U_{t}^{S} \equiv \mathbb{E}_{0}\sum_{t=0}^{\infty}\beta^{t}\frac{( C_{t}^{S}-\chi X_{t})^{1-\sigma}}{1-\sigma},
\end{equation}
subject to the budget constraint:
\begin{equation}
    W_{t}N_{t}^{S}+B_{t}+S_{t}( D_{t}+P_{t}^{s})+T_{t}^{S} = C_{t}^{S}+P_{t}^{b}B_{t+1}+P_{t}^{s}S_{t+1}. \label{eq:RicBC}
\end{equation}
The budget constraint states that consumption and purchases of equity shares (in quantity $S_{t+1}$ at price $P_{t}^{s}$) and risk-free bonds (in quantity $B_{t+1}$ at price $P_{t}^{b}$) must be financed by labor income $W_{t}N_{t}^{S}$ and the returns on financial investments. Shares purchased in the previous period yield a dividend $D_{t}$\todo{Are these dividends?}, while bonds yield a single consumption unit per bond in the following period. Moreover, the household receives lump-sum transfers from the government $T_{t}^{S}$ (or pays taxes, if negative).

\subsection{Government and market clearing}
The government runs a balanced budget, meaning that carbon tax revenues finance net transfers: 
\begin{equation}
    \tau_{t}E_{t} = T_{t}, \label{eq:GovBC}
\end{equation}
where aggregate lump-sum transfers are a weighted average of the two agents':
\begin{equation}
    T_{t} = \gamma T_{t}^{H}+(1- \gamma) T_{t}^{S}. \label{eq:AggT}
\end{equation}

Similarly, aggregate consumption and the equilibrium in the labor, stock, and bond markets are derived as:
\begin{eqnarray}
    C_{t} &=& \gamma C_{t}^{H}+(1- \gamma) C_{t}^{S}, \label{eq:AggC}
    \\
    N_{t} &=& \gamma N_{t}^{H}+(1- \gamma) N_{t}^{S},
    \\
    1 &=& (1-\gamma)S_{t},
    \\
    0 &=& (1-\gamma)B_{t},
\end{eqnarray}
where the equilibrium in financial markets accounts for limited asset market participation. While bonds are in zero net supply, the supply of stocks is normalized to $1$. Finally, market clearing in the goods market requires:
\begin{equation}
    Y_{t} = C_{t}+I_{t}+f(\mu_{t})Y_{t}, \label{eq:ResConstr}
\end{equation} 
i.e., total output is consumed, invested, or used to abate emissions.

\section{Competitive equilibrium and decentralized economy} \label{sec:Dec Economy}

We start by analyzing the competitive equilibrium, where firms maximize profits, and households maximize utility, taking the government's carbon tax rate as given. We show that the private marginal cost of emissions for firms is simply the carbon tax. Consequently, firms do not abate emissions in the decentralized economy (BAU scenario) where the tax rate is zero.

\paragraph{Households} Since they are excluded from financial markets and labor supply is inelastic, HtM households consume labor income every period. Conversely, the problem of the saver reads:

\begin{equation}
    \underset{\{ C_{t}^{S},B_{t+1},S_{t+1}\} }{\max } \quad U_{t}^{S} \equiv \mathbb{E}_{0}\sum_{t=0}^{\infty}\beta^{t}\frac{( C_{t}^{S}-\chi X_{t})^{1-\sigma}}{1-\sigma},
\end{equation}
subject to the budget constraint (\ref{eq:RicBC}). The associated Lagrangian is:
\begin{eqnarray*}
\mathcal{L}^{S} &\mathcal{=}&\mathbb{E}_{0}\sum_{t=0}^{\infty}\beta^{t}\Bigg\{\frac{(C_{t}^{S}-\chi X_{t})^{1-\sigma}}{(1-\sigma)}
\\
&& + \lambda_{t}^{S}[ W_{t}N_{t}^{S}+B_{t}+S_{t}(D_{t}+P_{t}^{s})+T_{t}^{S}-C_{t}^{S}-P_{t}^{b}B_{t+1}-P_{t}^{s}S_{t+1}] \Bigg\}.
\end{eqnarray*}
The resulting FOCs are:
\begin{eqnarray}
    C_{t}^{S} &:& \lambda_{t}^{S}=(C_{t}^{S}-\chi X_{t})^{-\sigma}, \label{eq:FOC_CR}
    \\
    B_{t+1} &:& P_{t}^{b}=\mathbb{E}_{t}\mathbb{M}_{t,t+1}^{S}, \label{eq:FOC_BR}
    \\
    S_{t+1} &:& P_{t}^{s}=\mathbb{E}_{t}\mathbb{M}_{t,t+1}^{S}( D_{t+1}+P_{t+1}^{s}). \label{eq:FOC_SR}
\end{eqnarray}
Equation (\ref{eq:FOC_CR}) denotes savers' marginal utility of consumption and $\mathbb{M}_{t,t+1}^{S}\equiv\beta(\lambda_{t+1}^{S}/\lambda_{t}^{S})$ the associated stochastic discount factor (SDF). Conditions (\ref{eq:FOC_BR})-(\ref{eq:FOC_SR}) are the key asset pricing equations for bonds and stocks. As usual, the bond price reflects expected marginal utility growth, while the stock price equals the expected discounted payoff, i.e., dividends plus the resale price.

\paragraph{Firms} The representative firm chooses output, labor, investment, capital stock, abatement, and emissions to maximize its value, using the saver (firm owner) marginal utility to discount payoffs:
\begin{equation}
    \underset{\{ Y_{t},N_{t},I_{t}.K_{t+1},\mu_{t},E_{t}\}}{\max} \quad \mathbb{E}_{0}\sum_{t=0}^{\infty}\beta^{t}\lambda_{t}^{S}D_{t},
\end{equation}
subject to the emission equation (\ref{eq:Em}), the production function (\ref{eq:ProdF}), the law of motion for capital (\ref{eq:KLoM}), and the adjustment cost function (\ref{eq:KAdjC}).

The associated Lagrangian is:
\begin{eqnarray*}
\mathcal{L}^{F}&\mathcal{=}&\mathbb{E}_{0}\sum_{t=0}^{\infty}\beta^{t}\lambda_{t}^{S}\Bigg\{Y_{t}-W_{t}N_{t}-I_{t}-f(\mu_{t})Y_{t}-E_{t}\tau_{t}
\\
&& + \varrho_{t}[\varepsilon_{t}^{A}AK_{t}^{\alpha }N_{t}^{1-\alpha}-Y_{t}]
\\
&& + Q_{t}\left[(1-\delta) K_{t}+\Phi\left(\frac{I_{t}}{K_{t}}\right)K_{t}-K_{t+1}\right]
\\
&& +V_{t}^{E}[E_{t}-\left( 1-\mu_{t}\right) \phi_{1}Y_{t}^{1-\phi_{2}}] \Bigg \}.
\end{eqnarray*}

The resulting first-order conditions are:
\begin{align}
    Y_{t} &: \varrho_{t}=1-f(\mu_{t})-V_{t}^{E}(1-\phi_{2}) E_{t}/Y_{t}, \label{eq:FOC_Y}
    \\
    N_{t} &: W_{t}=(1-\alpha) \varrho_{t}Y_{t}/N_{t}, \label{eq:FOC_N}
    \\
    I_{t} &: Q_{t}b_{1}\left(\frac{I_{t}}{K_{t}}\right)^{-\epsilon }=1, \label{eq:FOC_I}
    \\
    K_{t+1} &: Q_{t}=\mathbb{E}_{t}\mathbb{M}_{t,t+1}^{S}\Bigg\{ \varrho_{t+1}\alpha Y_{t+1}/K_{t+1}+Q_{t+1}\Bigg[(1-\delta) +\Phi \left(\frac{I_{t+1}}{K_{t+1}}\right)- \nonumber \\ 
    & \hspace{3.25cm} b_{1}\left(\frac{I_{t+1}}{K_{t+1}}\right)^{1-\epsilon}\Bigg]\Bigg\}, \label{eq:FOC_K}
    \\
    \mu_{t} &: V_{t}^{E}\frac{E_{t}}{1-\mu_{t}} =f^{'}(\mu)Y_{t}, \label{eq:FOC_MU}
    \\
    E_{t} &: V_{t}^{E}=\tau_{t}, \label{eq:FOC_E}
\end{align}
where $f(\mu_{t})^{'}=\theta_{1}\theta_{2}\mu_{t}^{\theta_{2}-1}$ is the marginal cost of abatement. 

Conditions (\ref{eq:FOC_N})-(\ref{eq:FOC_K}) are standard in a limited asset market participation setup, where the firm's SDF depends on savers' (rather than aggregate) marginal utility of consumption. Condition (\ref{eq:FOC_Y}) determines the real marginal cost of production. Producing one additional unit increases profits but the cost of abatement and the cost of higher emissions (the term $V_{t}^{E}(1-\phi_{2})E_{t}/Y_{t}$) need to be subtracted. Notice that $V_{t}^{E}$---the Lagrange multiplier on the emissions constraint--- can be interpreted as the firm's private marginal cost of increasing emissions (through larger production, as $(1-\phi_{2})E_{t}/Y_{t}=\partial E_{t}/\partial Y_{t}$). This is determined by condition (\ref{eq:FOC_E}) and equals the carbon tax rate. In a BAU scenario, the tax is not imposed. According to equation (\ref{eq:FOC_MU})---which governs the optimal level of abatement---if the tax is zero, then the firm finds it optimal not to abate ($\mu_t=0$), implying that the real marginal cost equals one, as in the standard RBC model.

Thus, in the competitive equilibrium firms do not consider the stock of GHGs to be a control variable and neglect their negative impact on household utility when making production decisions. Therefore, firms do not abate emissions without government intervention, with their shadow cost equal to zero. In the next section, we show that this generates inefficiency because the SCC---i.e., the implicit price of carbon for a social planner that maximizes welfare---is instead positive. Therefore, the optimal policy re-aligns the firm's private cost of emissions with the SCC.  

\section{Optimal carbon tax} \label{sec:Opt Tax}
This section solves the social planner's problem under different configurations. Specifically, we consider a utilitarian planner---meaning that the welfare weights are equal to the population shares---under different transfer policy constraints. In the unconstrained case, the planner can freely reallocate income across households. Then, the planner redistributes both tax revenues and financial income to eliminate consumption (and income) inequality between HtM and savers \citep[similar to][]{belfiori2024unequal}. In the constrained case, redistribution is restricted to tax revenues, resulting in consumption and income inequality in equilibrium.

\subsection{Unconstrained transfer policy}
Consider first the case where the planner can freely transfer income between the two groups of households. (The PU superscript will denote this case). The problem of the planner is to choose group-specific consumption, investment, capital stock, output, abatement, emissions, and the stock of GHGs to maximize welfare:
\begin{equation}
    \underset{\{ C_{t}^{H},C_{t}^{S},I_{t},K_{t+1},Y_{t},\mu_{t},E_{t},X_{t+1}\} }{\max} \quad \mathcal{W}_{t} = \gamma U_{t}^{H}+ (1-\gamma) U_{t}^{S}, 
\end{equation}
subject to the law of motion for GHGs (\ref{eq:GHGLoM}), the emission equation (\ref{eq:Em}), the production function (\ref{eq:ProdF}), the law of motion for capital (\ref{eq:KLoM}) and the associated adjustment cost function (\ref{eq:KAdjC}), the definition of aggregate consumption (\ref{eq:AggC}), and the resource constraint (\ref{eq:ResConstr}).

The associated Lagrangian is:
\begin{eqnarray*}
    \mathcal{L}^{PU} &\mathcal{=}&\mathbb{E}_{0}\sum_{t=0}^{\infty}\beta^{t}\Bigg\{\gamma \frac{(C_{t}^{H}-\chi X_{t})^{1-\sigma}}{(1-\sigma)} +(1-\gamma)\frac{(C_{t}^{S}-\chi X_{t})^{1-\sigma}}{(1-\sigma)}
    \\
    && + \lambda_{t}^{PU}[Y_{t}-\gamma C_{t}^{H}-(1-\gamma)C_{t}^{S}-I_{t}-f(\mu_{t})Y_{t}]
    \\
    && + \lambda_{t}^{PU} Q_{t}^{PU}\left[(1-\delta) K_{t}+\Phi\left(\frac{I_{t}}{K_{t}}\right) K_{t}-K_{t+1}\right]
    \\
    && + \lambda_{t}^{PU}\varrho_{t}^{PU}[\varepsilon_{t}^{A}AK_{t}^{\alpha} N_{t}^{1-\alpha }-Y_{t}]
    \\
    && + \lambda_{t}^{PU}V_{t}^{X,PU}[X_{t+1}-\eta X_{t}-E_{t}]
    \\
    && + \lambda_{t}^{PU}V_{t}^{E,PU}\left[E_{t}-(1-\mu_{t}) \phi_{1} Y_{t}^{1-\phi_{2}}\right]\Bigg\}.
\end{eqnarray*}

\paragraph{Consumption} The FOCs w.r.t. the two agents' consumption levels are: 
\begin{eqnarray}
    C_{t}^{H} &:& \lambda_{t}^{PU}=( C_{t}^{H}-\chi X_{t}) ^{-\sigma },
    \\
    C_{t}^{S} &:& \lambda_{t}^{PU}=( C_{t}^{S}-\chi X_{t}) ^{-\sigma },
\end{eqnarray}
which implies:
\begin{equation}
    C_{t}^{H}=C_{t}^{S}.
\end{equation}
If the planner can freely transfer resources between households, she\todo{could use they instead, but we need to be consistent} only needs to consider the aggregate---rather than individual---budget constraint. Therefore, the optimality condition requires consumption to be equalized across households, i.e., inequality is eliminated, and the resulting allocation resembles a representative agent economy. Recalling the expression for aggregate transfers---equation (\ref{eq:AggT})---it can be easily shown that equalized consumption entails:
\begin{eqnarray}
    T_{t}^{H} &=& \tau^{*}_{t}E_{t} + D_{t},
    \\
    T_{t}^{S} &=& \tau^{*}_{t}E_{t} - \frac{\gamma}{(1-\gamma)}D_{t},
\end{eqnarray}
that is, the planner uniformly redistributes the revenues from the optimal tax while transferring financial income from the saver to the HtM. 

\paragraph{Emissions, abatement, GHGs, and optimal tax level} Consider the FOCs wrt abatement, emissions, and the stock of GHGs in the atmosphere:
\begin{eqnarray}
    \mu_{t} &:& V_{t}^{E,PU}E_{t}/(1-\mu_{t})=f(\mu_{t})^{'}Y_{t},
    \\
    E_{t} &:& V_{t}^{E,PU}=V_{t}^{X,PU}, \label{eq:FOC_EU}
    \\
    X_{t+1} &:& V_{t}^{X,PU}=\mathbb{E}_{t}\mathbb{M}_{t,t+1}^{PU}[\eta V_{t+1}^{X,PU} +\chi],\label{eq:FOC_XU}
\end{eqnarray}
where $\mathbb{M}_{t,t+1}^{PU}\equiv\beta(\lambda_{t+1}^{PU}/\lambda_{t}^{PU})$ is the social planner's SDF.

The crucial difference with the decentralized economy is that condition (\ref{eq:FOC_EU}) equalizes the private cost of emissions to the \textit{social} cost of carbon. Therefore, the social planner internalizes the adverse effects of emissions because she considers the stock of GHGs to be a control variable. Comparing (\ref{eq:FOC_EU}) and (\ref{eq:FOC_E}), it follows that the optimal policy restores the first-best allocation in the competitive equilibrium by imposing the tax rate:
\begin{equation}
    \tau_{t}^{*}=V_{t}^{X,PU}, \label{eq:tau_U}
\end{equation}
i.e., by re-aligning the firm’s private cost of emission to the SCC. In turn, the policy induces firms to abate emissions optimally. The SCC---determined by condition (\ref{eq:FOC_XU})---is essentially an asset pricing formula. The social value of carbon equals the expected utility loss due to a unit increase in the stock of emissions ($\chi$), plus its continuation value---because emissions persist in the atmosphere according to the parameter $\eta$. The social planners' SDF, $\mathbb{M}_{t,t+1}^{PU}$, discounts this expected loss. As shown earlier, the latter is the same as the one that would prevail in a representative agent economy since consumption is equalized across agents through direct redistribution of financial income from the saver to the HtM household. 

\paragraph{Production and investment} Finally, the social planner also chooses output, investment, and capital stock:

\begin{align}
    Y_{t} &: \varrho_{t}^{PU} = 1-f\left( \mu_{t}\right)-V_{t}^{E,PU}(1-\phi_{2})E_{t}/Y_{t},
    \\
    I_{t} &: Q_{t}^{PU}b_{1}\left(\frac{I_{t}}{K_{t}}\right) ^{-\epsilon }=1,
    \\
    K_{t+1} &: Q_{t}^{PU}=\mathbb{E}_{t}\mathbb{M}_{t,t+1}^{PU}\Bigg\{ \varrho_{t+1}^{PU}\alpha Y_{t+1}/K_{t+1}+Q_{t+1}^{PU}\Bigg[(1-\delta) +\Phi \left(\frac{I_{t+1}}{K_{t+1}}\right)- \nonumber \\ 
    & \hspace{3.25cm} b_{1}\left(\frac{I_{t+1}}{K_{t+1}}\right)^{1-\epsilon}\Bigg]\Bigg\}. 
\end{align}
These optimality conditions are common to the RBC literature and are equivalent to the decentralized economy, except that the SDF depends on the planner's, rather than the saver's, marginal utility.

Next, we show that if the transfer policy is constrained to redistribute only the tax revenues (not also financial income), household heterogeneity modifies the SCC and the optimal tax.

\subsection{Constrained transfer policy}
Consider now the case where the social planner cannot \textit{directly} redistribute resources between households. Specifically, the planner can only transfer (lump-sum) the tax revenue to HtM and/or savers based on a policy of the form \citep[see, e.g.,][]{kanzig2023unequal}:
\begin{eqnarray}
    T_{t}^{S}&=&\frac{1-\xi}{1-\gamma}\tau_{t}E_{t}, \label{eq:TR_red}
    \\
    T_{t}^{H}&=&\frac{\xi}{\gamma}\tau_{t}E_{t}. \label{eq:TH_red}
\end{eqnarray}
The parameter $\xi\in[0,1]$ controls the share of revenues accruing to HtM households. For example, $\xi=0$ ($\xi=1$) implies that taxes are fully transferred to savers (HtM). If $\xi=\gamma$, instead, tax revenues are redistributed uniformly across all households. 

The problem of the planner reads:
\begin{equation}
    \underset{\{ C_{t}^{H},C_{t}^{S},I_{t},K_{t+1},Y_{t},\mu_{t},E_{t},X_{t+1}\} }{\max} \quad \mathcal{W}_{t} = \gamma U_{t}^{H}+ (1-\gamma) U_{t}^{S}, 
\end{equation}
subject to the same constraints as before, plus the individual HtM budget constraint in equation (\ref{eq:HtMBC}) and the transfer policy configuration (\ref{eq:TR_red})-(\ref{eq:TH_red}). Indeed, as direct redistribution is ruled out, the planner must consider individual budget constraints. However, given the aggregate resource constraint, the savers' budget constraint is redundant. (This case will be denoted by the superscript PC).

The associated Lagrangian is:
\begin{eqnarray*}
    \mathcal{L}^{PC} &\mathcal{=}&\mathbb{E}_{0}\sum_{t=0}^{\infty}\beta^{t}\Bigg\{\gamma \frac{(C_{t}^{H}-\chi X_{t})^{1-\sigma}}{(1-\sigma)} +(1-\gamma)\frac{(C_{t}^{S}-\chi X_{t})^{1-\sigma}}{(1-\sigma)}
    \\
    && + \lambda_{t}^{PC}[Y_{t}-\gamma C_{t}^{H}-(1-\gamma)C_{t}^{S}-I_{t}-f(\mu_{t})Y_{t}]
    \\
    && + \lambda_{t}^{PC}\lambda_{t}^{H,PC}[W_{t}N_{t}^{H}+\frac{\xi}{\gamma}\tau_{t}E_{t}- C_{t}^{H}]
    \\
    && + \lambda_{t}^{PC} Q_{t}^{PC}\left[(1-\delta) K_{t}+\Phi\left(\frac{I_{t}}{K_{t}}\right) K_{t}-K_{t+1}\right]
    \\
    && + \lambda_{t}^{PC}\varrho_{t}^{PC}[\varepsilon_{t}^{A}AK_{t}^{\alpha} N_{t}^{1-\alpha }-Y_{t}]
    \\
    && + \lambda_{t}^{PC}V_{t}^{X,PC}[X_{t+1}-\eta X_{t}-E_{t}]
    \\
    && + \lambda_{t}^{PC}V_{t}^{E,PC}\left[E_{t}-(1-\mu_{t}) \phi_{1} Y_{t}^{1-\phi_{2}}\right]\Bigg\}.
\end{eqnarray*}
Below, we discuss only the FOCs that differ from the unconstrained transfer policy scenario.

\paragraph{Consumption} The constrained problem gives rise to the following optimality conditions for consumption:
\begin{eqnarray}
    C_{t}^{H} &:& \lambda_{t}^{H,PC}=\gamma\frac{\left[(C_{t}^{H}-\chi X_{t})^{-\sigma}-\lambda_{t}^{PC}\right]}{\lambda_{t}^{PC}}, \label{eq:FOC_CHC}
    \\
    C_{t}^{S} &:& \lambda_{t}^{PC} =(C_{t}^{S}-\chi X_{t})^{-\sigma}. \label{eq:FOC_CRC}
\end{eqnarray}
Compared to the unconstrained case, the Lagrange multiplier on the resource constraint, $\lambda_{t}^{PC}$, is a function of savers' marginal utility. In other words, the planner's marginal utility coincides with the savers' one. The Lagrange multiplier on the HtM budget constraint, $\lambda_{t}^{H,PC}$, measures instead the relative difference between HtM and savers' marginal utility of consumption, weighted by the HtM population weight. Therefore, $\lambda_{t}^{H,PC}$ represents the key metric of inequality in the limited participation economy, which collapses to zero if asset market participation is full ($\gamma=0$).

\paragraph{Emissions, GHGs, and optimal tax level} The optimality conditions for emissions and the stock of GHGs now read:
\begin{eqnarray}
    E_{t} &:& V_{t}^{E,PC}=V_{t}^{X,PC}-\lambda_{t}^{H,PC}\frac{\xi}{\gamma}\tau_{t}, \label{eq:FOC_EC}
    \\
    X_{t+1} &:& V_{t}^{X,PC}=\mathbb{E}_{t}\mathbb{M}_{t,t+1}^{PC}\left\{\eta V_{t+1}^{X,PC}+\chi\left[1+\lambda_{t+1}^{H,PC}\right]\right\}, \label{eq:FOC_XC}
\end{eqnarray}
where $\mathbb{M}_{t,t+1}^{PC}\equiv\beta(\lambda_{t+1}^{PC}/\lambda_{t}^{PC})$ denotes the social planner's SDF in the constrained case. The SCC now features an additional term, $\chi\lambda_{t+1}^{H,PC}$, that accounts for the relatively higher disutility of GHGs related to the presence of poorer-than-average households (recall equation (\ref{eq:MargDis_X})).\footnote{As long as $\gamma\neq0$ and $C^H<C^R$, $\lambda_{t}^{H,PC}>0$. In other words, consumption inequality raises the average SCC relative to the unconstrained (representative agent) scenario.} Further, the social planner now equalizes the private cost of emissions ($V_{t}^{E,PC}$) to the SCC net of the marginal utility benefit coming from redistributing tax revenues towards HtM households (the term $\lambda_{t}^{H,PC}\frac{\xi}{\gamma}\tau_{t}$).

By imposing $V_{t}^{E,PC}=\tau_{t}^{*}$ in equation (\ref{eq:FOC_EC}), we can solve for the constrained optimal tax rate:
\begin{equation}
    \tau_{t}^{*}=\frac{V_{t}^{X,PC}}{1+\xi/\gamma\lambda_{t}^{H,PC}}. \label{eq:tau_C}
\end{equation}
Constraining the transfer policy available to the planner has two opposite effects on the optimal tax. On the one hand, emissions have a more negative impact on social welfare relative to the unconstrained case (i.e., $V_{t}^{X, PC}>V_{t}^{X,PU}$), because HtM households suffer relatively more from the accumulation of GHGs. The effect of the higher SCC calls for a higher tax. On the other hand, emissions generate revenues that the planner can use to close the consumption (income) gap between savers and HtM households. Recall that, in the first-best (unconstrained) policy, the planner finds it optimal to equalize consumption across agents. Thus, from the social planner's point of view, the redistributive benefit of positive emissions partially compensates for the negative externality due to the accumulation of GHGs. This effect calls for a lower tax rate, which would induce lower abatement efforts by producing firms. In this sense, an efficiency vs equality trade-off emerges in the presence of inequality and heterogeneous marginal disutility of pollution.

It is easy to see that the highest tax rate, and thus the highest abatement effort, is obtained when tax revenues are fully rebated to savers ($\xi=0$). If tax revenues are instead redistributed uniformly ($\xi=\gamma$), then the \emph{average} tax in the presence of inequality is aligned with the one prevailing in a representative agent economy. To see this, consider the steady-state SCC:
\begin{equation}
    V^{X,PC}=\frac{\beta}{1-\beta\eta}\chi\left[1+\lambda^{H,PC}\right],
\end{equation}
and the steady-state optimal tax:
\begin{equation}
    \tau^{*}=\frac{V^{X,PC}}{1+\xi/\gamma\lambda^{H,PC}}.
\end{equation}
It is easy to notice that if $\xi=\gamma$, then:
\begin{equation}
    \tau^{*}=\frac{\beta}{1-\beta\eta}\chi.
\end{equation}
Thus, the steady-state constrained optimal tax rate under uniform redistribution coincides with the unconstrained one (recall equation (\ref{eq:tau_U}), evaluated in the steady state). Finally, if the tax is disproportionately redistributed towards the HtM---i.e., if $\xi>\gamma$---then the tax is lower than the unconstrained one.

\paragraph{Optimal tax dynamics} Consumption and income inequality between savers and HtM households also have implications for the cyclical properties of the SCC and the optimal tax rate. To see this, suppose that the SCC is procyclical in the representative agent economy \citep[as found in][among the others]{heutel2012how}. Intuitively, during expansions, increased production generates higher emissions, which translates into a larger SCC---through the externality effect (captured by $\chi$) and the continuation value (captured by $\eta V_{t+1}^{X,PC}$). Thus, \emph{countercyclical} (\emph{procyclical}) consumption inequality will dampen (amplify) fluctuations in the SCC.\footnote{Notice that if consumption inequality is countercyclical, $\lambda_{t+1}^{H,PC}$ is countercyclical too.} Countercyclical consumption inequality means that, during expansions, the marginal utility of the HtM declines more relative to the savers’ one. In other words, poorer households enjoy a relatively more significant improvement in economic conditions that mitigates the negative social impact of the environmental externality, which is mainly concentrated among precisely these households. If inequality is strongly countercyclical, the positive effect of such a redistributive channel on HtMs' utility might even dominate, making the SCC potentially countercyclical too. In turn, the behavior of the SCC maps one-to-one into the dynamics of the optimal tax if the planner fully rebates the revenues to savers ($\xi=0$).

However, in the presence of redistribution ($\xi>0$), the dynamics of the marginal (dis)utility gap $\lambda_{t}^{H,PC}$ affects not only the SCC but also the planner's marginal benefit of redistribution, which enters in the denominator of the optimal tax. Thus, while dampening the dynamics of the SCC, countercyclical inequality could actually \textit{amplify} fluctuations in the optimal tax relative to the unconstrained case. This counteracting effect is more substantial for higher values of $\xi$, i.e. the larger the share redistributed to HtM households. Intuitively, suppose the planner is concerned about redistribution. In that case, she will find it optimal to bring the tax rate more in line with the SCC when inequality is low, that is, during expansions (if inequality is countercyclical). Even in the case of uniform redistribution ($\xi=\gamma$), which results in an \textit{average} rate that is equal to the representative agent economy, the decoupling between the SCC and tax rate dynamics holds, due to fluctuations in inequality that determine a time-varying wedge between the two variables. 

\section{Quantitative model} \label{sec:Quant Results}
In this section, we calibrate and simulate the model to provide a quantitative assessment of how the level and dynamics of inequality affect the behavior of the optimal tax along the business cycle. In line with the analytical intuition developed before, we find that the highest average tax rate is obtained in the constrained policy scenario with a rebate to the savers. In contrast, the tax rate becomes strongly procyclical and volatile if tax revenues are fully redistributed to HtM households. Indeed, in the baseline calibration inequality is markedly countercyclical, which implies that the planner's redistributive motive weakens during expansions. Overall, inequality plays a quantitatively relevant role in determining the SCC and the optimal carbon tax, both on average and along the business cycle.

\subsection{Calibration}
We calibrate the model to the U.S. economy at the quarterly frequency. The calibration is based on the BAU scenario because the U.S. has not implemented any carbon tax. Macroeconomic and environmental parameters follow \citet{heutel2012how}, while the parameters on habit preferences are taken from \citet{benmir2020green}. The resulting parameterization is summarized in Table \ref{tab:Calibration}.

We set the fraction of HtM households to $20\%$ to capture the lowest income quintile \citep[in line with][among the others]{eea2018unequal}. However, in Section \ref{sec:Robustness}, we consider alternative calibrations too---with $\gamma$ equal to $11\%$ or $33\%$, consistent with the literature distinguishing between "poor" and "wealthy" HtM \citep[see, e.g.,][]{kaplan2014wealthy,bilbiie2024monetary}. Households' discount rate is set to $\beta=0.98267$, whereas the local utility curvature and utility externality weight are $\sigma=4.199$ and $\chi=4\times10^{-4}$. In our calibration, the latter two parameter values imply that the agent-specific steady-state effective risk aversion ($RRA^{i}\equiv\frac{\sigma}{1-\chi X/C^{i}}$) ranges between $4.85$ and $5.98$ across the different policy scenarios, which lie well below the value of $10$ considered as an upper bound in the asset-pricing literature \citep{mehra1985equity}.\footnote{Specifically, the lowest values are obtained in the constrained optimal policy case with rebate to savers ($RRA^{S}=4.85$ and $RRA^{H}=5.11$). The highest ones emerge in the BAU scenario ($RRA^{S}=5.46$ and $RRA^{H}=5.98$).} As these parameters directly affect households' marginal disutility of pollution, we will discuss the sensitivity of our quantitative 
results to variations in $\sigma$ and $\chi$.

\begin{sidewaystable}[p] \centering 
    \caption{Baseline parameter values}
    \begin{tabular}{lP{3cm}L{3cm}c}
    \hline \hline		
    Description                      &    Parameter    	   &	Value   &	Source/Target   \\ 
    \multicolumn{4}{c}{} \\ \hline
    Fraction of HtM          &    $\gamma$	       & 	$0.20$   & Bottom income quintile       \\
    Discount rate            &    $\beta$	       &	$0.98267$  & \citet{heutel2012how}       \\
    Local utility curvature  &    $\sigma$		   &    $4.199$        & \citet{benmir2020green}       \\
    Externality weight       &    $\chi$		   &    $4\times10^{-4}$ & "  \\
    Labor supply             &    $N$		       &    $1$        & Inelastic labor supply  \\
    \multicolumn{4}{c}{} \\ 
    Steady-state TFP         &    $A$              &    $(N_{BAU}^{1-\alpha}K_{BAU}^{\alpha})^{-1}$     & $Y_{BAU}=1$ in steady state \\
    Capital share of income  &    $\alpha$	       & 	 $0.36$   & \citet{heutel2012how}       \\
    Depreciation rate        &    $\delta$	       &	 $0.025$  & "        \\
    Capital adjustment cost  &    $\epsilon$       &     $0.000$      & "        \\
    TFP persistence          &    $\rho^{A}$        &    $0.95$   & "       \\
    TFP shock volatility     &    $\sigma_{\eta^{A}}$ &  $0.007$   & "     \\
    \multicolumn{4}{c}{} \\ 
    Pollution decay          &    $\eta$	       & 	$0.9979$  & "        \\
    Abatement efficiency     &    $\theta_{1}$	   &	$0.05607$ & "       \\
    Abatement cost curvature &    $\theta_{2}$	   &	$2.8$     & "       \\
    Steady-state $E_{BAU}/Y_{BAU}$        &    $\phi_{1}$	   &	$1$ & "         \\
    Elasticity of $E_{t}$ to $Y_{t}$ &    $1-\phi_{2}$	   &	$0.696$ & "         \\
    \multicolumn{4}{c}{} \\ 
    Redistribution of tax revenues &    $\xi$	   &	$0 / \gamma / 1$ & To $S$/Uniform/To $H$        \\
    \hline \hline
    \end{tabular}
    \floatfoot{Notes: The model is simulated at a quarterly frequency.}
    \label{tab:Calibration}
\end{sidewaystable}

The production sector's parameters are standard in the RBC literature, with a capital share of income $\alpha=36\%$ and a capital depreciation rate $\delta=2.5\%$. In the baseline, we abstract from capital adjustment costs ($\epsilon=0$), although we test the sensitivity of our results to this parameter in Section \ref{sec:Robustness}. Regarding the exogenous technology process, TFP shocks are set to have volatility $\sigma_{\eta_{a}}=0.007$ and persistence $\rho^{a}=0.95$. The shifter $A$ is imposed to normalize output to $1$ only in the BAU steady state. This allows for welfare comparison between the BAU and optimal policy scenarios, as the tax affects the production level (and, thus, consumption) by changing firms' marginal costs.

Regarding environmental parameters, the pollution decay $\eta=0.9979$ implies that atmospheric carbon dioxide has a half-life of about $83$ years. Parameters $\theta_1$ and $\theta_2$ control the cost of abating emissions as a share of total output. The calibrated value of the $\theta_2$ ($2.8$) implies a convex cost function, while the coefficient $\theta_1=0.05067$ controls the average cost and, therefore, abatement efficiency. The latter dimension is particularly important for the planner's balance between efficiency and equity objectives, as shown in the robustness analysis. Furthermore, the parameters $\phi_1$ and $\phi_2$ govern the steady state emissions-to-output ratio and the elasticity of emissions to output, respectively. We set $\phi_1=1$ and $\phi_2=0.304$, implying an elasticity of $0.696$, as estimated in \citet{heutel2012how}.

Finally, three different transfer policy rules are contemplated for the constrained optimal policy analysis. Tax revenues are fully rebated to savers, uniformly redistributed to all households, or fully transferred to the HtM. These three alternatives map into $\xi=0 / \gamma / 1$.

\subsection{Optimal carbon tax and welfare implications}
Following \cite{KKSS08} we solve the model using a second-order approximation to the policy functions. Averages of key variables of interest are reported in Table \ref{tab:SimMeans}.\footnote{In the constrained policy scenario, the model's steady state needs to be found numerically. As the marginal utility gap enters the SCC but is also affected by the optimal tax rate, the SCC cannot be found analytically.} 

Consider first the BAU and unconstrained optimal policy (representative agent) scenarios. The imposition of the optimal tax reduces average output, consumption, and investment.\footnote{In the BAU scenario, average output is not exactly equal to $1$---as imposed in the non-stochastic steady state---due to the presence of aggregate risk that drives precautionary savings and, thus, a slightly higher average stock of capital.} A positive tax on emissions induces positive abatement ($\mu=32.2\%$), which raises abatement costs. Since the abatement technology is quite efficient in our baseline economy (i.e., $\theta_{1}$ is quite low), abatement costs still represent a small share of total output, despite the significant abatement effort. Indeed, in this economy, an optimal tax of $2\%$ \citep[the same value found in][]{benmir2020green} reduces emissions and the stock of GHGs by about $33\%$---from $1.001$ to $0.676$ and from $476.84$ to $321.79$, respectively---and the associated SCC. The stronger impact on emissions relative to consumption implies a substantial improvement in welfare ($\mathcal{W}_{t}$) which, however, is differently distributed between savers and HtM. Namely, the gain in aggregate welfare reflects the remarkable improvement in the poorer households' lifetime utility ($U_{t}^{H}$), who benefit the most from both the decline in the stock of GHGs and the rise in their consumption due to the redistribution (of both tax revenues and dividend income) that equalizes consumption across all households. Relative to the BAU scenario, also asset market participants' lifetime utility ($U_{t}^{S}$) slightly rises as the negative utility effect of lower consumption is dominated by the benefits of emissions abatement.

\begin{sidewaystable}[p] \centering 
    \caption{Means}
    \begin{tabular}{lP{3.5cm}P{0.5cm}P{3.5cm}P{0.5cm}P{3cm}P{3cm}P{3cm}}
    \hline \hline		
    Variable & BAU & & Unc. Opt. Policy & & \multicolumn{3}{c}{Con. Opt. Policy}   \\ \cmidrule{2-2} \cmidrule{4-4} \cmidrule{6-8}
             &     & &                      & &  Rebate to S ($\xi=0$) & Uniform Red. ($\xi=\gamma$) & Red. to HtM ($\xi=1$)   \\ \cmidrule{6-8}
    \multicolumn{4}{c}{} \\ \hline
    $Y_{t}$ & $1.002$ & & $0.996$ &  & $0.992$ & $0.995$ & $0.998$ \\
    \multicolumn{4}{c}{} \\ 
    $I_{t}$ & $0.212$ & & $0.208$ &  & $0.206$ & $0.208$ & $0.210$ \\
    $f(\mu_{t})Y_{t}$ & $0.000$ & & $0.002$ &  & $0.005$ & $0.002$ & $0.001$ \\
    $C_{t}$ & $0.790$ & & $0.785$ &  & $0.781$ & $0.785$ & $0.787$ \\
    \multicolumn{4}{c}{} \\ 
    $C^{S}_{t}$ & $0.827$ & & $0.785$ & & $0.820$ & $0.820$ & $0.816$ \\
    $C^{H}_{t}$ & $0.641$ & & $0.785$ & & $0.623$ & $0.643$ & $0.675$ \\
    \multicolumn{4}{c}{} \\ 
    $V^{X}_{t}$    & $0.034$ & & $0.020$ & & $0.032$ & $0.031$ & $0.028$ \\
    $\tau^{*}_{t}$ & $0.000$ & & $0.020$ & & $0.032$ & $0.020$ & $0.010$ \\
    $E_{t}$        & $1.001$ & & $0.676$ & & $0.580$ & $0.675$ & $0.780$ \\
    $\tau^{*}_{t}E_{t}$        & $0.000$ & & $0.014$ & & $0.019$ & $0.014$ & $0.008$ \\
    $X_{t}$        & $476.84$ & & $321.79$ & & $276.13$ & $321.42$ & $371.54$ \\
    $\mu_{t}$      & $0.000$ & & $0.322$ & & $0.417$ & $0.322$ & $0.218$ \\
    \multicolumn{4}{c}{} \\ 
    $\mathcal{W}_{t}$ & $-108.65$ & & $-69.895$ & & $-74.368$ & $-77.856$ & $-81.622$ \\
    $U_{t}^{S}$       & $-77.078$ & & $-69.895$ & & $-54.302$ & $-59.020$ & $-66.184$ \\
    $U_{t}^{H}$       & $-234.95$ & & $-69.895$ & & $-154.63$ & $-153.20$ & $-143.37$ \\
    \hline \hline
    \end{tabular}
    \floatfoot{Notes: Averages for selected macroeconomic and environmental variables, computed based on a second-order approximation to the policy functions. Statistics are reported for the business-as-usual (BAU), unconstrained optimal policy, and constrained optimal policy (under different redistribution schemes) scenarios.}
    \label{tab:SimMeans}
\end{sidewaystable}

Constraining the transfer policy only to redistribute tax revenues has quantitatively relevant effects on aggregate quantities, the optimal tax, and welfare. In line with the analytical intuition developed earlier, the highest (lowest) tax rate is obtained when tax revenues are fully rebated (redistributed) to savers (HtM). When $\xi=0$, the optimal tax rate equals $3.2\%$, i.e. $60\%$ higher than the unconstrained level. Therefore, for standard parameter values, consumption inequality---as captured by the marginal utility gap $\lambda_{t+1}^{H,PC}$---raises the SCC ($V^{X}_{t}$) in a quantitatively relevant manner (recall equation (\ref{eq:FOC_XC})). Intuitively, poorer households suffer greater disutility from the environmental externality, raising the SCC. The higher tax naturally induces more abatement and, thus, a larger drop in emissions and GHGs. However, the associated abatement costs imply lower output (through firms' marginal costs), investment, and consumption. Moreover, consumption inequality is exacerbated relative to the BAU scenario, with the consumption ratio $C_{t}^{S}/C_{t}^{H}$ rising from about $1.29$ to $1.32$. The latter effect is mostly due to a stronger decline in the HtM's consumption, reflecting a wage fall due to the higher marginal production costs generated by substantial abatement efforts. Despite this side effect, the policy is welfare-improving both at the aggregate and the agent-specific level, suggesting that, for the HtM, the cost of lower consumption and higher inequality are more than compensated by the benefit of strong emission reduction.

As shown in equation (\ref{eq:tau_C}), redistributing tax revenues to HtM households decouples the optimal tax rate from the SCC. As $\xi$ rises, both the SCC and the optimal tax decline, with the impact on the latter being more noticeable. Mechanically, as the tax rate drops, all aggregate quantities move toward the BAU economy (where $\tau=0$ by construction). Conversely, consumption inequality decreases because of two concurring channels. First, a lower tax raises wages through the firm's marginal cost. Second, although aggregate revenues fall, the net transfers received by poorer households become larger, further expanding their average consumption. In turn, this reduction in inequality compresses the SCC. This positive consumption effect is outweighed by the smaller cut in emissions, resulting in lower aggregate welfare compared to the unconstrained or tax rebate to S agents ($\xi=0)$ cases, although HtM households gain on net from the redistribution policy. Finally, as anticipated analytically, the tax under uniform redistribution coincides with the unconstrained one, generating almost identical aggregate and environmental implications. Nevertheless, as tax revenues alone are insufficient to eliminate inequality, poorer households with strong disutility from the environmental externality imply a higher SCC and more contained welfare improvements relative to the representative agent economy, especially for the HtM.

Overall, the results suggest that inequality is quantitatively relevant in determining the optimal tax level. Therefore, a relevant efficiency-equality trade-off emerges for the social planner. An unconstrained policy where a carbon tax and direct transfers across households are available achieves the first best, addressing both the environmental externality and consumption/income inequality. However, if only tax revenues are available for redistribution purposes, then the second-best policy calls for a high tax---reflecting a higher SCC due to the presence of poor households who particularly dislike the adverse externality effects---that is fully rebated to firms and thus firm owners. In other words, the planner finds it optimal to address the environmental externality even if this exacerbates inequality. Indeed, for poorer households, the benefit of emissions abatement dominates the negative effect of a (relatively small) reduction in their consumption. Nevertheless, as argued in Section \ref{sec:Robustness}, the design of the second-best policy critically hinges on the abatement technology efficiency parameter, $\theta_{1}$.

\subsection{Optimal carbon tax dynamics}

\begin{sidewaysfigure}[p] \centering
    \caption{Responses to TFP shocks}
    \includegraphics[width=\textwidth,height=11cm,trim=2cm 0.4cm 2.5cm 0.5cm,clip]{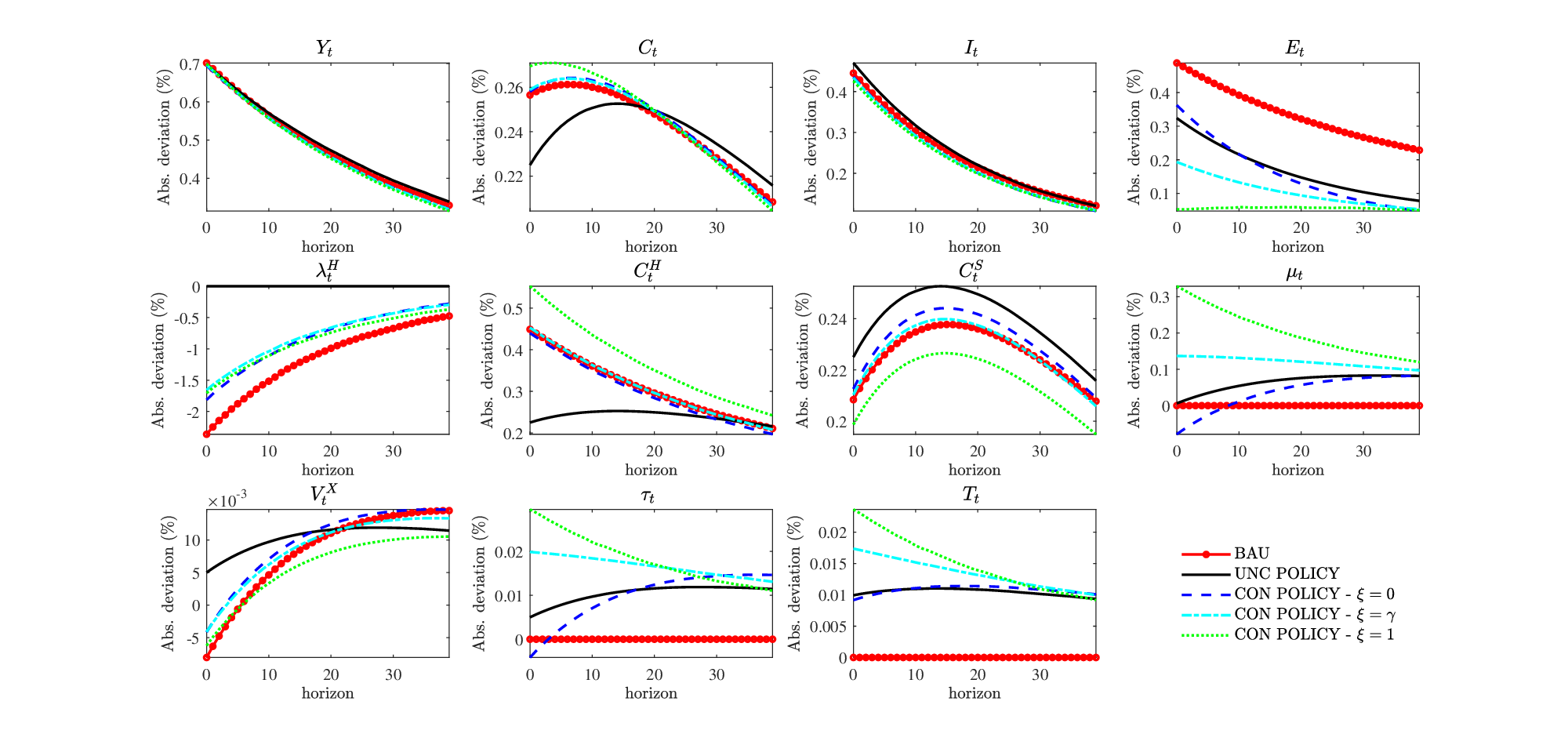}
    \vspace{-.7cm}
    \floatfoot{Notes: Impulse response functions (IRFs) of selected variables to a $1$ standard deviation (positive) TFP shock, computed based on a second-order approximation to the policy functions. IRFs are expressed in absolute deviations (multiplied by $100$) from the stochastic steady state and reported for the business-as-usual (BAU), unconstrained optimal policy, and constrained optimal policy (under different redistribution schemes) scenarios.}
    \label{fig:IRFs_TFP}
\end{sidewaysfigure}

Consumption and income inequality also affect the cyclical behavior of the optimal tax. To analyze the amplification/attenuation in the SCC and optimal tax dynamics due to time-varying inequality, Table \ref{tab:SimVola} reports model-implied standard deviations. Figure \ref{fig:IRFs_TFP} depicts the impulse response functions of selected variables to a one-standard-deviation positive TFP shock.

\paragraph{Standard deviations} The model under the BAU scenario performs similarly to standard RBC models, with (log) consumption (investment) being $0.70$ ($2.24$ times) as volatile as output. In the unconstrained (representative agent) scenario, aggregate volatility is slightly amplified, as the planner's redistribution policy stabilizes poorer households' consumption at the expense of the savers', directly affecting the standard deviation of investment---through savers' stochastic discount factor. Relatedly, the SCC exhibits a larger standard deviation relative to BAU, due to the removal of countercyclical fluctuations in the marginal utility gap ($\lambda_{t}^{H}$) that exert a dampening effect on the SCC's dynamics (as shown later). Naturally, introducing the tax reduces the variability in the emissions stream too. 

\begin{sidewaystable}[p] \centering 
    \caption{Standard deviations ($\%$)}
    \begin{tabular}{lP{3.5cm}P{0.5cm}P{3.5cm}P{0.5cm}P{3cm}P{3cm}P{3cm}}
    \hline \hline
    Variable & BAU & & Unc. Opt. Policy & & \multicolumn{3}{c}{Con. Opt. Policy} \\ \cmidrule{2-2} \cmidrule{4-4} \cmidrule{6-8} & & & & & Rebate to S ($\xi=0$) & Uniform Red. ($\xi=\gamma$) & Red. to HtM ($\xi=1$) \\\cmidrule{6-8} \multicolumn{4}{c}{} \\
    \hline
    $\log(Y_{t})$ & $3.62$ & & $3.68$ & & $3.55$ & $3.54$ & $3.51$ \\
    $\log(I_{t})$ & $8.12$ & & $8.53$ & & $7.89$ & $7.84$ & $7.66$ \\
    $\log(C_{t})$ & $2.54$ & & $2.55$ & & $2.52$ & $2.51$ & $2.50$ \\
    \multicolumn{4}{c}{} \\
    $\log(C^{S}_{t})$ & $2.35$ & & $2.55$ & & $2.35$ & $2.32$ & $2.22$ \\
    $\log(C^{H}_{t})$ & $3.62$ & & $2.55$ & & $3.52$ & $3.57$ & $4.03$ \\
    $\log(\lambda^{H}_{t})$ & $12.86$ & & $0.00$ & & $10.33$ & $11.53$ & $17.85$ \\
    \multicolumn{4}{c}{} \\
    $\log(\tau^{*}_{t})$ & $0.00$ & & $4.86$ & & $3.76$ & $6.37$ & $13.06$ \\
    $\log(E_{t})$ & $2.52$ & & $1.77$ & & $1.99$ & $1.12$ & $0.66$ \\
    $\log(V^{X}_{t})$ & $3.73$ & & $4.86$ & & $3.76$ & $3.65$ & $3.22$ \\
    \hline \hline
    \end{tabular}
    \floatfoot{Notes: Standard deviations (in $\%$) for selected macroeconomic and environmental variables, computed based on a second-order approximation to the policy functions. Statistics are reported for the business-as-usual (BAU), unconstrained optimal policy, and constrained optimal policy (under different redistribution schemes) scenarios.}
    \label{tab:SimVola}
\end{sidewaystable} 

The constrained optimal policy has a limited stabilizing impact on macroeconomic aggregates compared to the BAU benchmark. However, the different redistributive schemes significantly alter the behavior of agent-specific consumption, inequality, SCC, and optimal tax. When moving from the rebate ($\xi=0$) to the full redistribution ($\xi=1$) policy, the marginal utility gap dynamics are strongly amplified as the volatility of the HtMs' consumption rises relative to the savers' one. In turn, as anticipated analytically, larger fluctuations in the marginal utility gap \textit{dampen} the variations in SCC while \textit{amplifying} those in the tax rate, which map into relatively stable emission flows. Recalling equation (\ref{eq:tau_C}), these results suggest that the countercyclical time-variation in the planner's redistributive motive (the denominator) dominates that in SCC (the numerator), resulting in a strong amplification of the optimal tax dynamics as fiscal revenues are redistributed towards poorer households.

\paragraph{Impulse response analysis} To provide intuition about the model-implied dynamics, Figure \ref{fig:IRFs_TFP} displays the responses of selected variables to a positive 1-standard deviation TFP shock. To account for the fact that different scenarios are approximated around different steady states (as shown in Table \ref{tab:SimMeans}), we report \textit{absolute} deviations (multiplied by $100$) from the respective stochastic steady state. The BAU economy (red-dots solid line) features standard business-cycle dynamics, with higher TFP raising output, consumption, and investment. Expanded production raises emissions, whereas the absence of a carbon tax induces no abatement by polluting firms and leaves transfers to zero. Concurrently, savers' consumption rises substantially less than the HtMs', which maps into markedly countercyclical consumption inequality---as captured by the negative response of the marginal (dis)utility gap ($\lambda_{t}^{H}$). Such a response drives a fall in the SCC, which declines despite the rise in emissions. Therefore, the redistributive effect of the shock, which mitigates the aggregate marginal disutility of pollution by favoring HtM consumption during expansions, dominates the rise in the stock of GHGs in the determination of the SCC.

The countercyclical behavior of the SCC is a novel result relative to existing literature that abstracts from household heterogeneity in the study of optimal carbon tax dynamics. In this respect, it is instructive to compare the responses under BAU with those under the unconstrained optimal policy---which results in a representative agent economy (black solid lines). From an aggregate perspective, while unchanged for output, the emissions response is substantially smaller and less persistent relative to BAU, due to the hump-shaped response of the optimal tax inducing a gradual increase in abatement effort over time. Therefore, as in \citet{heutel2012how} and \citet{benmir2020green}, the SCC---and, thus, the optimal tax---is \textit{procyclical} in the representative agent economy, where the marginal utility gap is zero by definition.

The countercyclical behavior of the SCC also holds in the constrained optimal policy scenarios, as the marginal (dis)utility gap always declines in the limited participation economy. However, as argued analytically, only if tax revenues are fully rebated to savers does the tax rate exactly reflect the dynamics of the SCC. Hence, when $\xi=0$ (blue dashed lines), the tax rate slightly falls on impact to then climb above average only after about $5$ periods, producing a steeper profile of the emissions positive response. From a redistributive perspective, the expansion in emissions entails that tax revenues (and transfers) still rise, and the consequent rebate to savers slightly mitigates the fall in consumption inequality relative to BAU. When instead revenues are partially ($\xi=\gamma$) or fully ($\xi=1$) redistributed to the HtM (cyan dash-dotted and green dotted lines, respectively), the dynamics of the tax rate strongly diverge from that of the SCC. Specifically, despite the SCC being always countercyclical, the tax rate becomes strongly \textit{procyclical}. In these cases, therefore, abatement swiftly reacts to the positive technology shock, effectively containing the surge in emissions. Moreover, full transfers to HtM households sustain their consumption relative to asset market participants. 

Why does the tax rate behave so differently in the presence of redistribution? Recall that when only tax revenues are available to address consumption inequality, the optimal tax (\ref{eq:tau_C}) reflects an efficiency vs equality trade-off. On the one hand, the planner would like to set a high tax that significantly curbs the stock of GHGs, thus addressing the environmental externality. On the other hand, emissions generate tax revenues that can be used to close the consumption gap between asset market participants and HtM households, with this redistributive motive becoming stronger the larger the share of revenues to redistribute to the poor. As reported in Table \ref{tab:SimMeans}, this tension is resolved in favor of the redistributive objective, with the \textit{average} tax rate declining as $\xi$ rises. However, the trade-off evolves \textit{dynamically} along the business cycle. In our calibrated economy, consumption inequality shrinks following a positive technology shock. Therefore, during expansions, the planner's redistributive motive becomes weaker relative to the climate objective, triggering a stronger hike in the tax rate. In other words, the planner finds it optimal to take advantage of expansions---i.e. times of lower inequality---to decisively address the negative effects of the climate externality, resulting in a much more procyclical tax compared to the representative agent case.

Overall, in a similar vein to \citet{kanzig2023unequal}, we find that the constrained optimal policy with redistribution stabilizes both emissions and inequality, which is lower on average and exhibits smaller absolute deviations from the steady state relative to BAU. However, recall that IRFs are expressed in absolute deviations from different stochastic steady states. Therefore, a similar absolute deviation in inequality (or the tax rate) between different redistribution schemes entails very different responses \textit{relative} to the steady state. For example, in the full redistribution ($\xi=1$) policy, inequality is on average lower than in the full rebate ($\xi=0$) case (recall Table \ref{tab:SimMeans}). Thus, in relative terms, the inequality dynamics are strongly amplified as $\xi$ rises, consistent with the information conveyed by the standard deviations discussed earlier.

\section{Sensitivity analysis} \label{sec:Robustness}

In this section, we analyze the robustness of the relevance of cyclical inequality for both the level and the dynamics of the SCC and the optimal carbon tax along some key parameter values. The resulting tables and figures are reported in Appendix \ref{app:Appendix A}.

\paragraph{Different HtM shares} As a first robustness exercise, we consider a lower or higher share of HtM households by setting the parameter $\gamma=$ to $0.11$ or $0.33$. These shares are in line with \citet{kaplan2014wealthy}, who show that about one third of US households are HtM, with only a third of them (hence, about $11\%$ of total population) being "poor" in terms of illiquid wealth. Tables \ref{tab:SimMeans_LowGamma}-\ref{tab:SimVola_HighGamma}, and Figures \ref{fig:IRFs_TFP_LowGamma}-\ref{fig:IRFs_TFP_HighGamma}, demonstrate that, as expected, a lower (higher) share of HtM weakens (strengthens) the quantitative implications of inequality for the SCC and the optimal tax. However, the qualitative insights remain unaltered. 

\paragraph{Low abatement efficiency} Our results underline a trade-off between efficiency and equity for the design of the optimal carbon tax. In this perspective, the parameter $\theta_{1}$---which controls the abatement technology efficiency through the abatement cost function---can, in principle, change the second-best policy, which, in the baseline analysis, consisted of a high tax (inducing large abatement efforts) with rebate to savers. Thus, we simulate the model with $\theta_{1}$ being $3.5$ times the baseline. Such a value implies that the planner can induce half as much abatement in the unconstrained policy scenario given the \textit{same tax rate}.\footnote{Since inequality affects the optimal tax, in the constrained scenario it would not be possible to induce less abatement without changing also the optimal tax rate. This is possible in the representative agent economy, where the average SCC is a function of $\beta$, $\eta$, and $\chi$.} Table \ref{tab:SimMeans_HighTheta1} confirms this conjecture. As the abatement technology is less efficient, the planner needs a tax that is too high to improve the HtM's lifetime utility through reduced emissions significantly. Consequently, the second and third-best policies imply redistributing revenues  to the HtM or uniformly, respectively. Therefore, the prevalence of efficiency vs equity objectives crucially hinges on the planner's ability to curb emissions through the carbon tax. Table \ref{tab:SimVola_HighTheta1} and Figure \ref{fig:IRFs_TFP_HighTheta1} show instead that this parameter is almost irrelevant to the model dynamics.

\paragraph{Utility function parameters} The habit utility function parameters directly affect households' risk aversion, elasticity of intertemporal substitution, and valuation of climate damages. Thus, we repeat the analysis by halving the local curvature parameter ($\sigma=2$) or the surplus consumption ratio $\left(1-\chi X/C^i\right)$ in the BAU non-stochastic steady state (by setting $\chi=8.7360\times10^{-4}$).\footnote{Recall that half $\sigma$ (surplus consumption) maps into half (double) the baseline effective risk aversion in the BAU scenario. The resulting values still lie towards the lower bound usually considered in the external habit literature \citep[see][for comparison]{benmir2020green}.} Tables \ref{tab:SimMeans_LowSigma}-\ref{tab:SimVola_HighChi} and Figures \ref{fig:IRFs_TFP_LowSigma}-\ref{fig:IRFs_TFP_HighChi} show that these parameters affect both the means and the standard deviations of all variables. Lower $\sigma$ (higher $\chi$) implies larger (smaller) marginal disutility from emissions, resulting in lower (higher) average SCC and optimal tax rates in the scenarios with inequality.\footnote{The average unconstrained optimal tax and SCC are not affected by $\sigma$ instead.} Similarly, lower $\sigma$ (higher $\chi$) dampen (amplify) the dynamics of the marginal (dis)utility gap, implying that the SCC becomes more procyclical (countercyclical). In all cases, the relationship between inequality and optimal taxes remains intact from both a qualitative and quantitative point of view.

\paragraph{Capital adjustment costs} Finally, we consider the sensitivity of our results to the capital adjustment cost parameter by setting $\epsilon=1.5$ \citep[similar to][]{benmir2020green}. By construction, this parameter does not affect the steady state but only the model dynamics (Tables \ref{tab:SimMeans_HighEps}-\ref{tab:SimVola_HighEps} and Figure \ref{fig:IRFs_TFP_HighEps}). By making investment less volatile and, thus, dividends less countercyclical, capital adjustment costs make the marginal (dis)utility gap less countercyclical. As a result, the SCC and optimal tax rate are always procyclical (even in the unequal economy under BAU and constrained optimal policy). However, the dampening (amplifying) effect of countercyclical inequality on the SCC (tax) dynamics remains quantitatively relevant, as suggested by the change in the standard deviations moving from the unconstrained scenario to the constrained optimal policy with a full redistribution to HtM households. 

\section{Conclusion} \label{sec:Conclusions}
This paper explores the implications of household heterogeneity for the design and cyclical behavior of optimal carbon taxes in a real economy. Given our focus on balancing the potentially heterogeneous costs and benefits of a carbon tax, we incorporate the environmental externality directly into the utility function. By allowing for complementarity between consumption and the stock of emissions, our setup introduces inequality in the marginal disutility of pollution, disproportionately affecting poorer households. The presence of heterogeneous agents thus shapes both the level and dynamics of the optimal carbon tax, revealing an efficiency-equity trade-off for the social planner.

Starting from a benchmark decentralized setup, in the absence of a carbon tax (BAU scenario) firms have no incentive to internalize the environmental externality, leading to zero abatement in equilibrium. We then consider, moving to a centralized solution, two distinct scenarios. First, the unconstrained case, where the planner can redistribute income freely across households, effectively eliminating consumption and income inequality. Here, the optimal carbon tax aligns with the SCC, reflecting the first-best solution that addresses environmental externality without being influenced by household heterogeneity.

In the constrained case, where redistribution is limited to tax revenues, key differences arise. The social cost of carbon reflects the immediate and future loss of utility from emissions and incorporates the increased disutility experienced by poorer households. Moreover, the planner faces a trade-off: higher emissions reduce welfare due to their environmental impact but also generate additional tax revenues that can be used to alleviate and stabilize inequality. This trade-off results in an optimal tax rate influenced by environmental and redistributive considerations. Our analysis highlights that, under constrained redistribution, inequality amplifies the SCC and leads to higher average tax rates. However, if revenues are redistributed to poorer households, the tax rate decreases relative to the unconstrained case, and its fluctuations become more pronounced in the presence of countercyclical inequality.

The second key contribution of this paper lies in examining the cyclical properties of the SCC and the optimal carbon tax. We show that countercyclical consumption inequality mitigates fluctuations in the SCC as poorer households disproportionately benefit from economic expansions, which reduces the social impact of emissions externalities. Contrary to findings in the literature, this effect can even render the SCC countercyclical under strongly countercyclical inequality. Furthermore, the relationship between the SCC and the optimal tax is contingent on redistribution policies. While the tax directly tracks the SCC when revenues are fully rebated to savers, redistribution introduces a time-varying wedge between the two, amplifying the cyclical variability of the tax rate, particularly when redistribution is targeted toward poorer households.

Finally, we quantitatively assess how the level and dynamics of inequality influence the optimal carbon tax along the business cycle. The results confirm that inequality plays a quantitatively significant role in shaping both the average level and the cyclical behavior of the tax. Notably, the findings underscore the planner's efficiency and equity dilemma. In the unconstrained scenario, the planner achieves the first-best solution by addressing both environmental externalities and inequality. In the constrained case, the second-best policy calls for a higher tax to mitigate the disproportionate environmental impact on poorer households, even at the cost of exacerbating inequality. However, this approach ultimately benefits poorer households by prioritizing emission reduction over a small reduction in their consumption.

These findings have important implications for climate policy, emphasizing the need to consider household heterogeneity and the efficiency-equity trade-off we highlight when designing carbon taxation policies. Future research could extend this framework to explore the role of endogenous labor supply, nominal rigidities, and relaxation of the government-balanced budget assumption in shaping the optimal carbon tax. Furthermore, our setup can also be used to analyze optimal carbon taxation under different Pareto weights (e.g., to introduce an inequality-averse planner) or to study the dynamics in response to additional shocks (e.g., to emissions from the rest of the world or to the carbon tax itself). 


\bibliography{bibliography.bib}
\bibliographystyle{apalike} 

\newpage

\newgeometry{top=1in,left=1in,right=1in}
\linespread{1.25}\selectfont

\emptythanks
\setcounter{footnote}{1}

\begin{titlepage}
\title{\textit{Internet Appendix to}\\ Optimal Carbon Tax with \\ Limited Asset Market Participation}
\maketitle
\thispagestyle{empty}
\end{titlepage}

\begin{appendices}

\section{Sensitivity analysis} \label{app:Appendix A}
    \numberwithin{figure}{section}
    \numberwithin{table}{section}
    \setcounter{figure}{0}
    \setcounter{table}{0}

\begin{sidewaystable}[p] \centering 
    \caption{Means - Poor HtM}
    \begin{tabular}{lP{3.5cm}P{0.5cm}P{3.5cm}P{0.5cm}P{3cm}P{3cm}P{3cm}} \hline \hline 
    Variable & BAU & & Unc. Opt. Policy & & \multicolumn{3}{c}{Con. Opt. Policy} \\ \cmidrule{2-2} \cmidrule{4-4} \cmidrule{6-8} 
    & & & & & Rebate to S ($\xi=0$) & Uniform Red. ($\xi=\gamma$) & Red. to HtM ($\xi=1$) \\\cmidrule{6-8} \multicolumn{4}{c}{} \\ \hline 
    $Y_{t}$ & $1.002$ & & $0.996$ & & $0.994$ & $0.995$ & $0.997$ \\ \multicolumn{4}{c}{} \\ 
    $I_{t}$ & $0.212$ & & $0.208$ & & $0.207$ & $0.208$ & $0.209$ \\ $f(\mu_{t})Y_{t}$ & $0.000$ & & $0.002$ & & $0.003$ & $0.002$ & $0.001$ \\ 
    $C_{t}$ & $0.790$ & & $0.785$ & & $0.783$ & $0.785$ & $0.787$ \\
    \multicolumn{4}{c}{} \\ 
    $C^{S}_{t}$ & $0.808$ & & $0.785$ & & $0.802$ & $0.802$ & $0.794$ \\ $C^{H}_{t}$ & $0.642$ & & $0.785$ & & $0.626$ & $0.643$ & $0.726$ \\ 
    \multicolumn{4}{c}{} \\ 
    $V^{X}_{t}$ & $0.027$ & & $0.020$ & & $0.026$ & $0.025$ & $0.022$ \\ 
    $\tau^{*}_{t}$ & $0.000$ & & $0.020$ & & $0.026$ & $0.020$ & $0.014$ \\ 
    $E_{t}$ & $1.001$ & & $0.676$ & & $0.628$ & $0.675$ & $0.739$ \\ 
    $\tau^{*}_{t}E_{t}$ & $0.000$ & & $0.014$ & & $0.016$ & $0.014$ & $0.010$ \\ 
    $X_{t}$ & $476.91$ & &$321.79$ & & $299.10$ & $321.58$ & $352.14$ \\ 
    $\mu_{t}$ & $0.000$ & & $0.322$ & & $0.369$ & $0.322$ & $0.258$ \\ 
    \multicolumn{4}{c}{} \\ 
    $\mathcal{W}_{t}$ & $-101.37$ & & $-69.895$ & & $-72.427$ & $-74.006$ & $-74.369$ \\ 
    $U_{t}^{S}$ & $-84.863$ & & $-69.895$ & & $-61.520$ & $-64.216$ & $-70.803$ \\ 
    $U_{t}^{H}$ & $-234.92$ & & $-69.895$ & & $-160.67$ & $-153.22$ & $-103.22$ \\ 
    \hline \hline 
    \end{tabular}
    \floatfoot{Notes: Averages for selected macroeconomic and environmental variables with $\gamma=0.11$.}
    \label{tab:SimMeans_LowGamma}
\end{sidewaystable}

\begin{sidewaystable}[p] \centering 
    \caption{Standard deviations ($\%$) - Poor HtM}
    \begin{tabular}{lP{3.5cm}P{0.5cm}P{3.5cm}P{0.5cm}P{3cm}P{3cm}P{3cm}}
    \hline \hline		
    Variable & BAU & & Unc. Opt. Policy & & \multicolumn{3}{c}{Con. Opt. Policy}   \\ \cmidrule{2-2} \cmidrule{4-4} \cmidrule{6-8}
             &     & &                      & &  Rebate to S ($\xi=0$) & Uniform Red. ($\xi=\gamma$) & Red. to HtM ($\xi=1$)   \\\cmidrule{6-8}
    \multicolumn{4}{c}{} \\ \hline
    $\log(Y_{t})$ & 	$3.68$ & & $3.68$ & & $3.61$ & $3.61$ & $3.50$ \\
    $\log(I_{t})$ & 	$8.43$ & & $8.53$ & & $8.19$ & $8.16$ & $7.65$ \\
    $\log(C_{t})$ & 	$2.55$ & & $2.55$ & & $2.53$ & $2.53$ & $2.48$ \\
    \multicolumn{4}{c}{} \\ 
    $\log(C^{S}_{t})$   & $2.45$ & & $2.55$ & &  $2.44$ & $2.43$ & $2.20$ \\
    $\log(C^{H}_{t})$   & $3.68$ & & $2.55$ & & $3.58$ & $3.63$ & $5.20$ \\
    $\log(\lambda^{H}_{t})$ & $13.03$ & & $0.00$ & & $10.67$ & $11.69$ & $44.57$ \\
    \multicolumn{4}{c}{} \\ 
    $\log(\tau^{*}_{t})$ & $0.00$ & & $4.86$ & & $4.18$ & $5.69$ & $19.20$ \\
    $\log(E_{t})$ & 	$2.56$ & & $1.77$ & & $1.88$ & $1.39$ & $1.15$ \\
    $\log(V^{X}_{t})$ & $4.30$ & & $4.86$ & & $4.18$ & $4.15$ & $3.42$ \\
    \hline \hline
\end{tabular}

    \floatfoot{Notes: Standard deviations for selected macroeconomic and environmental variables with $\gamma=0.11$.}
    \label{tab:SimVola_LowGamma}
\end{sidewaystable}  

\begin{sidewaysfigure}[p] \centering
    \caption{Responses to TFP shocks - Poor HtM}
    \includegraphics[width=\textwidth,height=11cm,trim=2cm 0.4cm 2.5cm 0.5cm,clip]{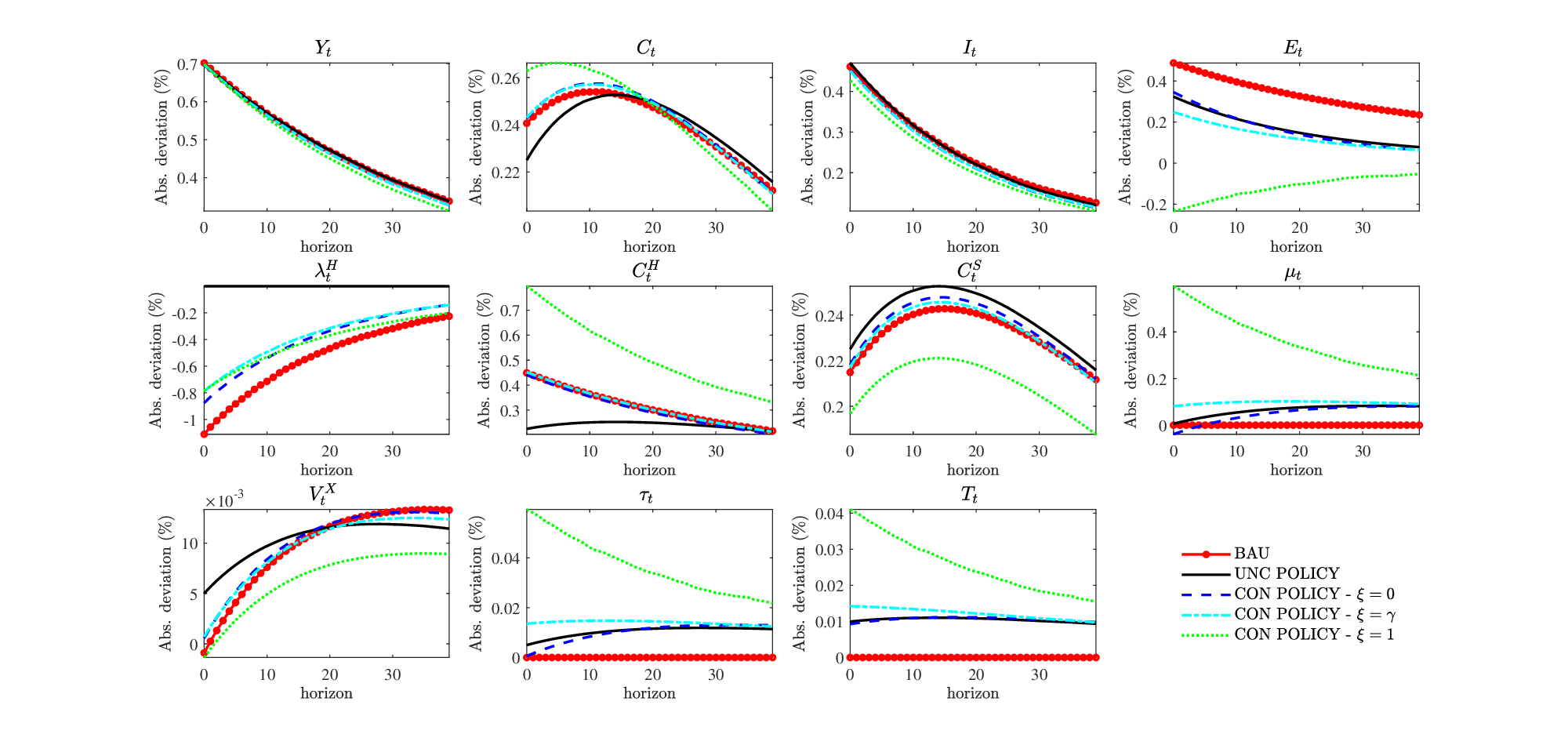}
    \vspace{-.7cm}
    \floatfoot{Notes: Impulse response functions (IRFs) of selected variables to a $1$ standard deviation (positive) TFP shock with $\gamma=0.11$.}
    \label{fig:IRFs_TFP_LowGamma}
\end{sidewaysfigure}

\begin{sidewaystable}[p] \centering 
    \caption{Means - Poor and wealthy HtM}
    \begin{tabular}{lP{3.5cm}P{0.5cm}P{3.5cm}P{0.5cm}P{3cm}P{3cm}P{3cm}}
    \hline \hline		
    Variable & BAU & & Unc. Opt. Policy & & \multicolumn{3}{c}{Con. Opt. Policy}   \\ \cmidrule{2-2} \cmidrule{4-4} \cmidrule{6-8}
             &     & &                      & &  Rebate to S ($\xi=0$) & Uniform Red. ($\xi=\gamma$) & Red. to HtM ($\xi=1$)   \\\cmidrule{6-8}
    \multicolumn{4}{c}{} \\ \hline
    $Y_{t}$ & 	$1.002$ & & $0.996$ &  & $0.988$ & $0.995$ & $0.998$ \\
    \multicolumn{4}{c}{} \\ 
    $I_{t}$ & 	$0.212$ & & $0.208$ &  & $0.204$ & $0.208$ & $0.210$ \\
    $f(\mu_{t})Y_{t}$ & $0.000$ & & $0.002$ &  & $0.008$ & $0.002$ & $0.001$ \\
    $C_{t}$ & 	$0.790$ & & $0.785$ &  & $0.776$ & $0.785$ & $0.787$ \\
    \multicolumn{4}{c}{} \\ 
    $C^{S}_{t}$ & $0.863$ & & $0.785$ & & $0.854$ & $0.854$ & $0.850$ \\
    $C^{H}_{t}$ & $0.641$ & & $0.785$ & & $0.617$ & $0.643$ & $0.659$ \\
    \multicolumn{4}{c}{} \\ 
    $V^{X}_{t}$    & $0.050$ & & $0.020$ & & $0.046$ & $0.042$ & $0.040$ \\
    $\tau^{*}_{t}$ & $0.000$ & & $0.020$ & & $0.046$ & $0.020$ & $0.010$ \\
    $E_{t}$        & $1.001$ & & $0.676$ & & $0.488$ & $0.675$ & $0.777$ \\
    $\tau^{*}_{t}E_{t}$        & $0.000$ & & $0.014$ & & $0.023$ & $0.014$ & $0.008$ \\
    $X_{t}$        & $476.74$ & &$321.79$ & & $232.61$ & $321.19$ & $370.09$ \\
    $\mu_{t}$      & $0.000$ & & $0.322$ & & $0.507$ & $0.323$ & $0.221$ \\
    \multicolumn{4}{c}{} \\ 
    $\mathcal{W}_{t}$ & $-120.81$ & & $-69.895$ & & $-76.698$ & $-84.384$ & $-89.286$ \\
    $U_{t}^{S}$       & $-64.579$ & & $-69.895$ & & $-43.433$ & $-50.503$ & $-56.064$ \\
    $U_{t}^{H}$       & $-234.98$ & & $-69.895$ & & $-144.24$ & $-153.17$ & $-156.74$ \\
    \hline \hline
    \end{tabular}
    \floatfoot{Notes: Averages for selected macroeconomic and environmental variables with $\gamma=0.33$.}
    \label{tab:SimMeans_HighGamma}
\end{sidewaystable}

\begin{sidewaystable}[p] \centering 
    \caption{Standard deviations ($\%$) - Poor and wealthy HtM}
    \begin{tabular}{lP{3.5cm}P{0.5cm}P{3.5cm}P{0.5cm}P{3cm}P{3cm}P{3cm}}
    \hline \hline		
    Variable & BAU & & Unc. Opt. Policy & & \multicolumn{3}{c}{Con. Opt. Policy}   \\ \cmidrule{2-2} \cmidrule{4-4} \cmidrule{6-8}
             &     & &                      & &  Rebate to S ($\xi=0$) & Uniform Red. ($\xi=\gamma$) & Red. to HtM ($\xi=1$)   \\\cmidrule{6-8}
    \multicolumn{4}{c}{} \\ \hline
    $\log(Y_{t})$ & 	$3.51$ & & $3.68$ & & $3.45$ & $3.44$ & $3.42$ \\
    $\log(I_{t})$ & 	$7.63$ & & $8.53$ & & $7.44$ & $7.34$ & $7.23$ \\
    $\log(C_{t})$ & 	$2.51$ & & $2.55$ & & $2.50$ & $2.48$ & $2.49$ \\
    \multicolumn{4}{c}{} \\ 
    $\log(C^{S}_{t})$   & $2.18$ & & $2.55$ & &  $2.20$ & $2.15$ & $2.08$ \\
    $\log(C^{H}_{t})$   & $3.51$ & & $2.55$ & & $3.43$ & $3.47$ & $3.68$ \\
    $\log(\lambda^{H}_{t})$ & $12.62$ & & $0.00$ & & $9.78$ & $11.29$ & $14.06$ \\
    \multicolumn{4}{c}{} \\ 
    $\log(\tau^{*}_{t})$ & $0.00$ & & $4.86$ & & $3.32$ & $7.26$ & $11.14$ \\
    $\log(E_{t})$ & 	$2.45$ & & $1.77$ & & $2.29$ & $0.77$ & $0.83$ \\
    $\log(V^{X}_{t})$ & $3.20$ & & $4.86$ & & $3.32$ & $3.08$ & $2.81$ \\
    \hline \hline
    \end{tabular}
    \floatfoot{Notes: Standard deviations for selected macroeconomic and environmental variables with $\gamma=0.33$.}
    \label{tab:SimVola_HighGamma}
\end{sidewaystable}  

\begin{sidewaysfigure}[p] \centering
    \caption{Responses to TFP shocks - Poor and wealthy HtM}
    \includegraphics[width=\textwidth,height=11cm,trim=2cm 0.4cm 2.5cm 0.5cm,clip]{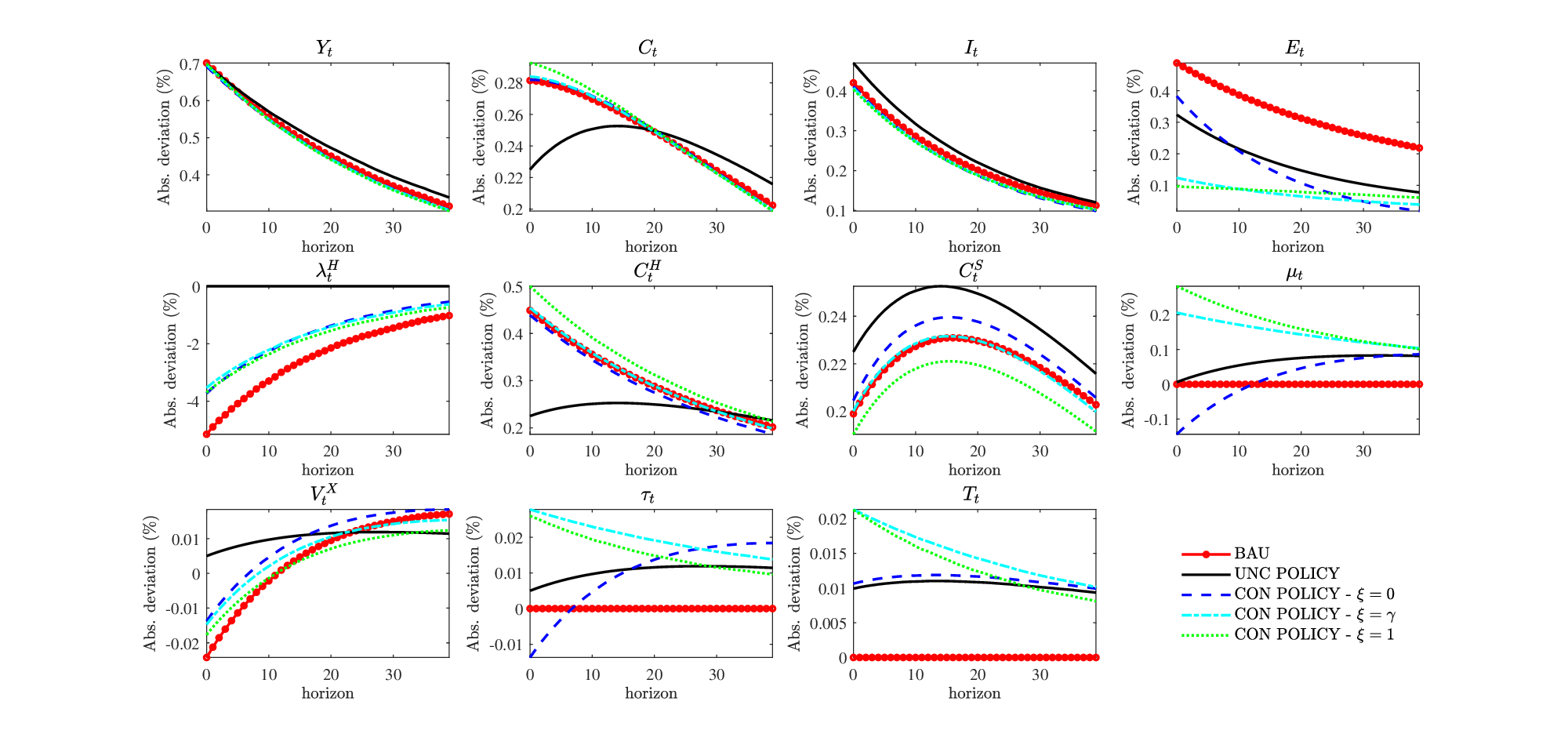}
    \vspace{-.7cm}
    \floatfoot{Notes: Impulse response functions (IRFs) of selected variables to a $1$ standard deviation (positive) TFP shock with $\gamma=0.33$.}
    \label{fig:IRFs_TFP_HighGamma}
\end{sidewaysfigure}

\begin{sidewaystable}[p] \centering 
    \caption{Means - Low abatement efficiency}
    \begin{tabular}{lP{3.5cm}P{0.5cm}P{3.5cm}P{0.5cm}P{3cm}P{3cm}P{3cm}} \hline \hline 
    Variable & BAU & & Unc. Opt. Policy & & \multicolumn{3}{c}{Con. Opt. Policy} \\ \cmidrule{2-2} \cmidrule{4-4} \cmidrule{6-8}
    & & & & & Rebate to S ($\xi=0$) & Uniform Red. ($\xi=\gamma$) & Red. to HtM ($\xi=1$) \\ \cmidrule{6-8} 
    \multicolumn{4}{c}{} \\ \hline 
    $Y_{t}$ & $1.002$ & & $0.995$ & & $0.990$ & $0.995$ & $0.998$ \\ \multicolumn{4}{c}{} \\ 
    $I_{t}$ & $0.212$ & & $0.208$ & & $0.205$ & $0.208$ & $0.210$ \\ 
    $f(\mu_{t})Y_{t}$ & $0.000$ & & $0.001$ & & $0.003$ & $0.001$ & $0.000$ \\ 
    $C_{t}$ & $0.790$ & & $0.786$ & & $0.782$ & $0.786$ & $0.788$ \\ 
    \multicolumn{4}{c}{} \\ 
    $C^{S}_{t}$ & $0.827$ & & $0.786$ & & $0.823$ & $0.821$ & $0.815$ \\ 
    $C^{H}_{t}$ & $0.641$ & & $0.786$ & & $0.620$ & $0.645$ & $0.680$ \\ 
    \multicolumn{4}{c}{} \\ 
    $V^{X}_{t}$ & $0.034$ & & $0.020$ & & $0.035$ & $0.031$ & $0.027$ \\ 
    $tau^{*}_{t}$ & $0.000$ & & $0.020$ & & $0.035$ & $0.020$ & $0.010$ \\ 
    $E_{t}$ & $1.001$ & & $0.836$ & & $0.778$ & $0.836$ & $0.889$ \\ 
    $\tau^{*}_{t}E_{t}$ & $0.000$ & & $0.017$ & & $0.027$ & $0.017$ & $0.009$ \\ 
    $X_{t}$ & $476.84$ & & $398.27$ & & $370.40$ & $397.97$ & $423.23$ \\ 
    $\mu_{t}$ & $0.000$ & & $0.161$ & & $0.216$ & $0.161$ & $0.110$ \\ 
    \multicolumn{4}{c}{} \\ 
    $\mathcal{W}_{t}$ & $-108.65$ & & $-81.078$ & & $-91.70$ & $-91.245$ & $-90.418$ \\ 
    $U_{t}^{S}$ & $-77.078$ & & $-81.078$ & & $-63.939$ & $-68.044$ & $-73.593$ \\ 
    $U_{t}^{H}$ & $-234.95$ & & $-81.078$ & & $-202.74$ & $-184.05$ & $-157.72$ \\ 
    \hline \hline 
    \end{tabular}
    \floatfoot{Notes: Averages for selected macroeconomic and environmental variables with $\theta_{1}=0.05607\times 3.5$.}
    \label{tab:SimMeans_HighTheta1}
\end{sidewaystable}
    
\begin{sidewaystable}[p] \centering 
    \caption{Standard deviations ($\%$) - Low abatement efficiency}
    \begin{tabular}{lP{3.5cm}P{0.5cm}P{3.5cm}P{0.5cm}P{3cm}P{3cm}P{3cm}} 
    \hline \hline 
    Variable & BAU & & Unc. Opt. Policy & & \multicolumn{3}{c}{Con. Opt. Policy} \\ \cmidrule{2-2} \cmidrule{4-4} \cmidrule{6-8} 
    & & & & & Rebate to S ($\xi=0$) & Uniform Red. ($\xi=\gamma$) & Red. to HtM ($\xi=1$) \\ \cmidrule{6-8} 
    \multicolumn{4}{c}{} \\ \hline 
    $\log(Y_{t})$ & $3.62$ & & $3.72$ & & $3.58$ & $3.57$ & $3.51$ \\ 
    $\log(I_{t})$ & $8.12$ & & $8.66$ & & $8.02$ & $7.95$ & $7.66$ \\ 
    $\log(C_{t})$ & $2.54$ & & $2.57$ & & $2.54$ & $2.53$ & $2.51$ \\ 
    \multicolumn{4}{c}{} \\ 
    $\log(C^{S}_{t})$ & $2.35$ & & $2.57$ & & $2.37$ & $2.33$ & $2.18$ \\ 
    $\log(C^{H}_{t})$ & $3.62$ & & $2.57$ & & $3.55$ & $3.64$ & $4.27$ \\ 
    $\log(\lambda^{H}_{t})$ & $12.86$ & & $0.00$ & & $11.04$ & $12.57$ & $21.18$ \\ 
    \multicolumn{4}{c}{} \\ 
    $\log(\tau^{*}_{t})$ & $0.00$ & & $5.08$ & & $3.87$ & $6.85$ & $14.77$ \\ 
    $\log(E_{t})$ & $2.52$ & & $2.23$ & & $2.25$ & $1.88$ & $1.53$ \\ 
    $\log(V^{X}_{t})$ & $3.73$ & & $5.08$ & & $3.87$ & $3.66$ & $3.05$ \\ 
    \hline \hline 
    \end{tabular}
    \floatfoot{Notes: Standard deviations for selected macroeconomic and environmental variables with $\theta_{1}=0.05607\times 3.5$.}
    \label{tab:SimVola_HighTheta1}
\end{sidewaystable} 

\begin{sidewaysfigure}[p] \centering
    \caption{Responses to TFP shocks - Low abatement efficiency}
    \includegraphics[width=\textwidth,height=11cm,trim=2cm 0.4cm 2.5cm 0.5cm,clip]{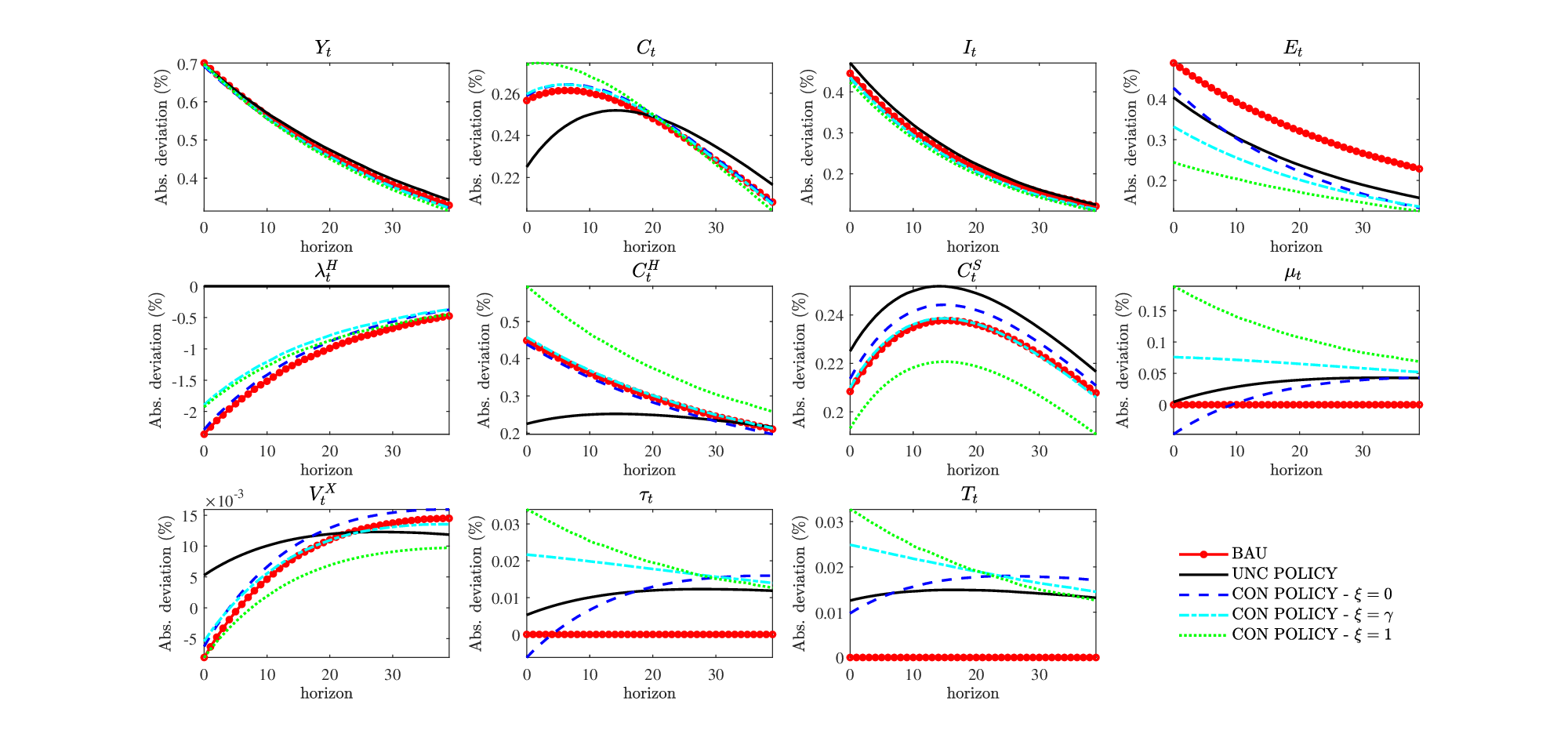}
    \vspace{-.7cm}
    \floatfoot{Notes: Impulse response functions (IRFs) of selected variables to a $1$ standard deviation (positive) TFP shock with $\theta_{1}=0.05607\times 3.5$.}
    \label{fig:IRFs_TFP_HighTheta1}
\end{sidewaysfigure}

\begin{sidewaystable}[p] \centering 
    \caption{Means - Low risk aversion}
    \begin{tabular}{lP{3.5cm}P{0.5cm}P{3.5cm}P{0.5cm}P{3cm}P{3cm}P{3cm}}
    \hline \hline		
    Variable & BAU & & Unc. Opt. Policy & & \multicolumn{3}{c}{Con. Opt. Policy}   \\ \cmidrule{2-2} \cmidrule{4-4} \cmidrule{6-8}
             &     & &                      & &  Rebate to S ($\xi=0$) & Uniform Red. ($\xi=\gamma$) & Red. to HtM ($\xi=1$)   \\ \cmidrule{6-8}
    \multicolumn{4}{c}{} \\ \hline
    $Y_{t}$ & $1.001$ & & $0.994$ &  & $0.993$ & $0.994$ & $0.996$ \\
    \multicolumn{4}{c}{} \\ 
    $I_{t}$ & $0.212$ & & $0.208$ &  & $0.207$ & $0.208$ & $0.209$ \\
    $f(\mu_{t})Y_{t}$ & $0.000$ & & $0.002$ &  & $0.003$ & $0.002$ & $0.001$ \\
    $C_{t}$ & $0.789$ & & $0.784$ &  & $0.783$ & $0.784$ & $0.786$ \\
    \multicolumn{4}{c}{} \\ 
    $C^{S}_{t}$ & $0.827$ & & $0.784$ & & $0.822$ & $0.820$ & $0.811$ \\
    $C^{H}_{t}$ & $0.641$ & & $0.784$ & & $0.627$ & $0.642$ & $0.686$ \\
    \multicolumn{4}{c}{} \\ 
    $V^{X}_{t}$    & $0.024$ & & $0.020$ & & $0.024$ & $0.024$ & $0.022$ \\
    $\tau^{*}_{t}$ & $0.000$ & & $0.020$ & & $0.024$ & $0.020$ & $0.015$ \\
    $\tau^{*}_{t}E_{t}$        & $0.000$ & & $0.014$ & & $0.015$ & $0.014$ & $0.011$ \\
    $E_{t}$        & $1.001$ & & $0.676$ & & $0.643$ & $0.676$ & $0.728$ \\
    $X_{t}$        & $476.45$ & & $321.85$ & & $306.43$ & $321.80$ & $346.72$ \\
    $\mu_{t}$      & $0.000$ & & $0.321$ & & $0.353$ & $0.321$ & $0.270$ \\
    \multicolumn{4}{c}{} \\ 
    $\mathcal{W}_{t}$ & $-98.339$ & & $-88.096$ & & $-88.958$ & $-89.357$ & $-89.860$ \\
    $U_{t}^{S}$       & $-85.917$ & & $-88.096$ & & $-77.764$ & $-79.595$ & $-81.735$ \\
    $U_{t}^{H}$       & $-128.49$ & & $-88.096$ & & $-114.64$ & $-112.52$ & $-105.71$ \\
    \hline \hline
    \end{tabular}
    \floatfoot{Notes: Averages for selected macroeconomic and environmental variables with $\sigma=2$.}
    \label{tab:SimMeans_LowSigma}
\end{sidewaystable}
    
\begin{sidewaystable}[p] \centering 
    \caption{Standard deviations ($\%$) -  Low risk aversion}
    \begin{tabular}{lP{3.5cm}P{0.5cm}P{3.5cm}P{0.5cm}P{3cm}P{3cm}P{3cm}}
    \hline \hline
    Variable & BAU & & Unc. Opt. Policy & & \multicolumn{3}{c}{Con. Opt. Policy} \\ \cmidrule{2-2} \cmidrule{4-4} \cmidrule{6-8} & & & & & Rebate to S ($\xi=0$) & Uniform Red. ($\xi=\gamma$) & Red. to HtM ($\xi=1$) \\\cmidrule{6-8} \multicolumn{4}{c}{} \\
    \hline
    $\log(Y_{t})$ & $3.29$ & & $3.34$ & & $3.26$ & $3.25$ & $3.24$ \\
    $\log(I_{t})$ & $6.93$ & & $7.32$ & & $6.82$ & $6.80$ & $6.68$ \\
    $\log(C_{t})$ & $2.50$ & & $2.52$ & & $2.50$ & $2.50$ & $2.49$ \\
    \multicolumn{4}{c}{} \\
    $\log(C^{S}_{t})$ & $2.37$ & & $2.52$ & & $2.39$ & $2.38$ & $2.31$ \\
    $\log(C^{H}_{t})$ & $3.29$ & & $2.52$ & & $3.24$ & $3.26$ & $3.48$ \\
    $\log(\lambda^{H}_{t})$ & $7.98$ & & $0.000$ & & $6.79$ & $7.40$ & $11.49$ \\
    \multicolumn{4}{c}{} \\
    $\log(\tau^{*}_{t})$ & $0.00$ & & $2.77$ & & $2.54$ & $3.07$ & $5.42$ \\
    $\log(E_{t})$ & $2.29$ & & $1.98$ & & $1.99$ & $1.77$ & $1.35$ \\
    $\log(V^{X}_{t})$ & $2.68$ & & $2.77$ & & $2.54$ & $2.53$ & $2.45$ \\
    \hline \hline
    \end{tabular}
    \floatfoot{Notes: Standard deviations for selected macroeconomic and environmental variables with $\sigma=2$.}
    \label{tab:SimVola_LowSigma}
\end{sidewaystable} 

\begin{sidewaysfigure}[p] \centering
    \caption{Responses to TFP shocks - Low risk aversion}
    \includegraphics[width=\textwidth,height=11cm,trim=2cm 0.4cm 2.5cm 0.5cm,clip]{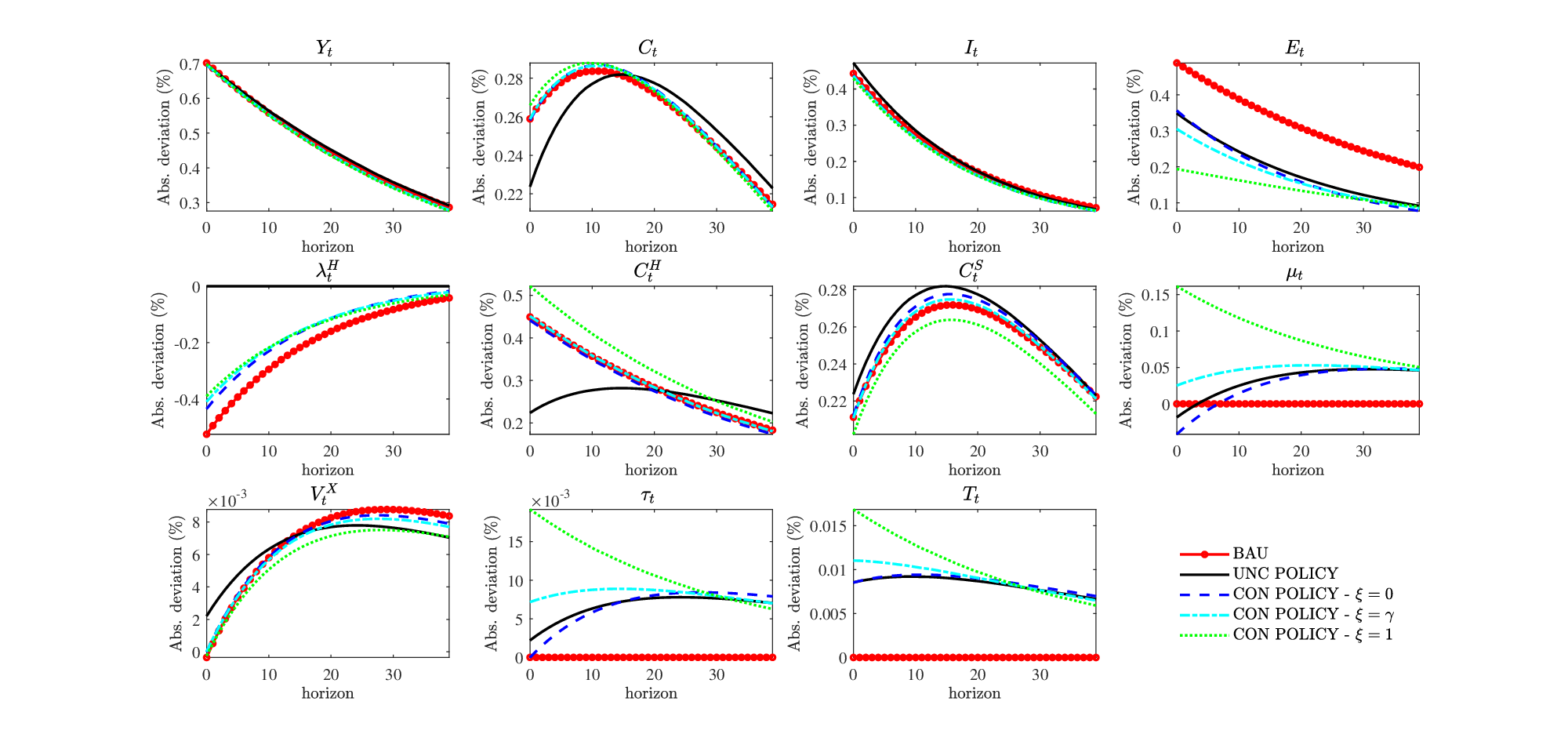}
    \vspace{-.7cm}
    \floatfoot{Notes: Impulse response functions (IRFs) of selected variables to a $1$ standard deviation (positive) TFP shock with $\sigma=2$.}
    \label{fig:IRFs_TFP_LowSigma}
\end{sidewaysfigure}

\begin{sidewaystable}[p] \centering 
    \caption{Means - High externality weight}
    \begin{tabular}{lP{3.5cm}P{0.5cm}P{3.5cm}P{0.5cm}P{3cm}P{3cm}P{3cm}}
    \hline \hline
    Variable & BAU & & Unc. Opt. Policy & & \multicolumn{3}{c}{Con. Opt. Policy} \\ \cmidrule{2-2} \cmidrule{4-4} \cmidrule{6-8} & & & & & Rebate to S ($\xi=0$) & Uniform Red. ($\xi=\gamma$) & Red. to HtM ($\xi=1$) \\\cmidrule{6-8} \multicolumn{4}{c}{} \\
    \hline
    $Y_{t}$ & $1.004$ & & $0.989$ & & $0.982$ & $0.989$ & $0.994$ \\
    $I_{t}$ & $0.213$ & & $0.205$ & & $0.201$ & $0.205$ & $0.207$ \\
    $f(\mu_{t})Y_{t}$ & $0.000$ & & $0.008$ & & $0.018$ & $0.008$ & $0.003$ \\
    $C_{t}$ & $0.790$ & & $0.777$ & & $0.764$ & $0.776$ & $0.783$ \\
    \multicolumn{4}{c}{} \\
    $C^{S}_{t}$ & $0.827$ & & $0.777$ & & $0.803$ & $0.810$ & $0.802$ \\
    $C^{H}_{t}$ & $0.642$ & & $0.777$ & & $0.606$ & $0.640$ & $0.707$ \\
    \multicolumn{4}{c}{} \\
    $V^{X}_{t}$ & $0.156$ & & $0.045$ & & $0.075$ & $0.072$ & $0.057$ \\
    $\tau^{*}_{t}$ & $0.000$ & & $0.045$ & & $0.075$ & $0.045$ & $0.025$ \\
    $E_{t}$ & $1.002$ & & $0.498$ & & $0.331$ & $0.497$ & $0.636$ \\
    $\tau^{*}_{t}E_{t}$ & $0.000$ & & $0.022$ & & $0.025$ & $0.022$ & $0.016$ \\
    $X_{t}$ & $477.37$ & & $237.26$ & & $157.48$ & $236.47$ & $302.84$ \\
    $\mu_{t}$ & $0.000$ & & $0.498$ & & $0.665$ & $0.499$ & $0.360$ \\
    \multicolumn{4}{c}{} \\
    $\mathcal{W}_{t}$ & $-709.19$ & & $-110.21$ & & $-94.71$ & $-126.24$ & $-156.96$ \\
    $U_{t}^{S}$ & $-315.55$ & & $-110.21$ & & $-66.69$ & $-91.18$ & $-132.10$ \\
    $U_{t}^{H}$ & $-2283.80$ & & $-110.21$ & & $-206.80$ & $-266.50$ & $-256.39$ \\
    \hline \hline
    \end{tabular}
    \floatfoot{Notes: Averages for selected macroeconomic and environmental variables with $\chi=8.7360 \times 10^{-4}$.}
    \label{tab:SimMeans_HighChi}
\end{sidewaystable}
    
\begin{sidewaystable}[p] \centering 
    \caption{Standard deviations ($\%$) - High externality weight}
    \begin{tabular}{lP{3.5cm}P{0.5cm}P{3.5cm}P{0.5cm}P{3cm}P{3cm}P{3cm}}
    \hline \hline
    Variable & BAU & & Unc. Opt. Policy & & \multicolumn{3}{c}{Con. Opt. Policy} \\ \cmidrule{2-2} \cmidrule{4-4} \cmidrule{6-8} & & & & & Rebate to S ($\xi=0$) & Uniform Red. ($\xi=\gamma$) & Red. to HtM ($\xi=1$) \\\cmidrule{6-8} \multicolumn{4}{c}{} \\
    \hline
    $\log(Y_{t})$ & $3.94$ & & $3.75$ & & $3.57$ & $3.56$ & $3.42$ \\
    $\log(I_{t})$ & $9.27$ & & $8.77$ & & $8.03$ & $7.89$ & $7.18$ \\
    $\log(C_{t})$ & $2.61$ & & $2.53$ & & $2.48$ & $2.45$ & $2.39$ \\
    \multicolumn{4}{c}{} \\
    $\log(C^{S}_{t})$ & $2.38$ & & $2.53$ & & $2.30$ & $2.25$ & $1.91$ \\
    $\log(C^{H}_{t})$ & $3.94$ & & $2.53$ & & $3.53$ & $3.54$ & $4.75$ \\
    $\log(\lambda^{H}_{t})$ & $29.82$ & & $0.000$ & & $11.12$ & $13.77$ & $36.85$ \\
    \multicolumn{4}{c}{} \\
    $\log(\tau^{*}_{t})$ & $0.00$ & & $5.27$ & & $3.58$ & $7.32$ & $21.42$ \\
    $\log(E_{t})$ & $2.74$ & & $1.30$ & & $2.71$ & $1.01$ & $3.81$ \\
    $\log(V^{X}_{t})$ & $5.98$ & & $5.27$ & & $3.58$ & $3.36$ & $2.27$ \\
    \hline \hline
    \end{tabular}
    \floatfoot{Notes: Standard deviations for selected macroeconomic and environmental variables with $\chi=8.7360 \times 10^{-4}$.}
    \label{tab:SimVola_HighChi}
\end{sidewaystable} 

\begin{sidewaysfigure}[p] \centering
    \caption{Responses to TFP shocks - High externality weight}
    \includegraphics[width=\textwidth,height=11cm,trim=2cm 0.4cm 2.5cm 0.5cm,clip]{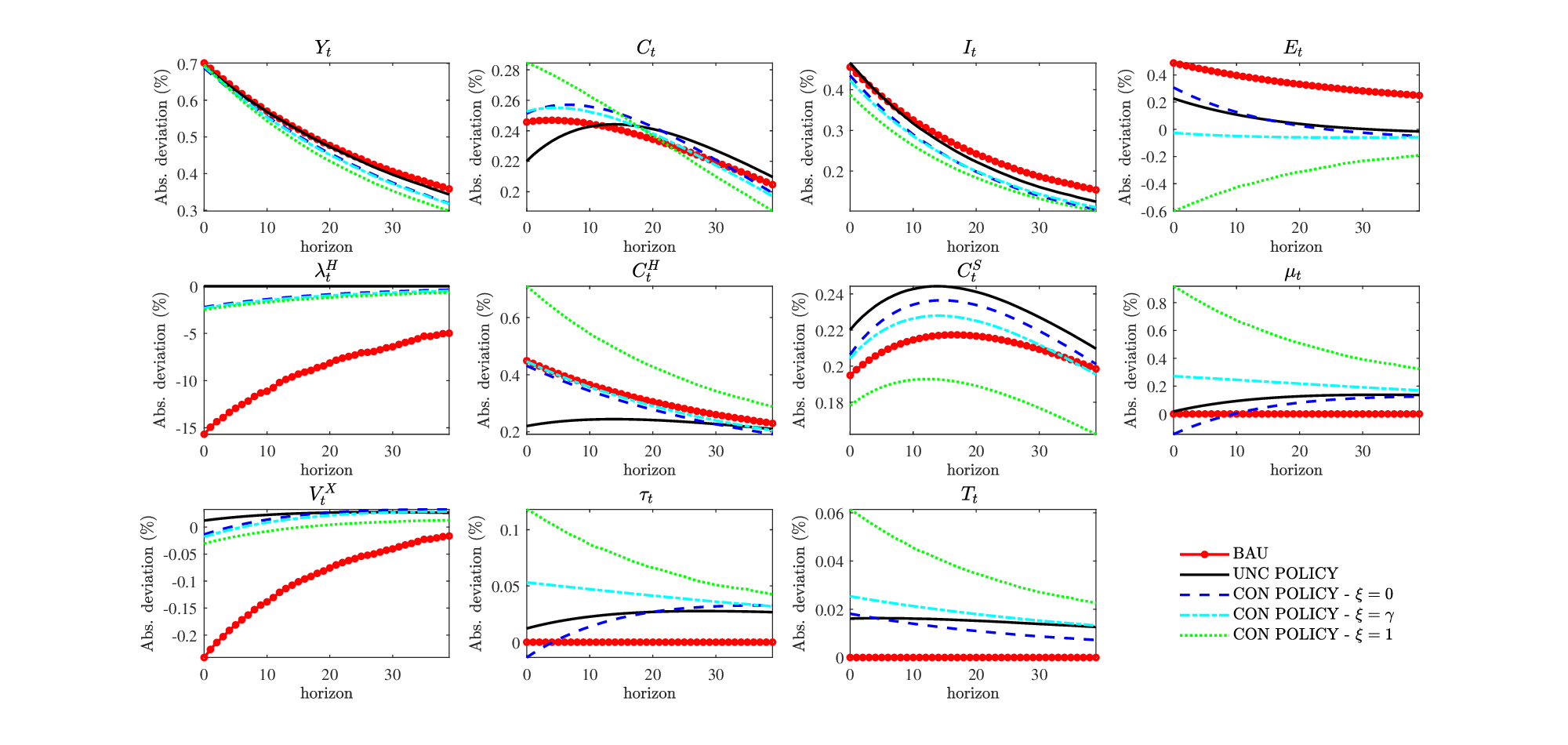}
    \vspace{-.7cm}
    \floatfoot{Notes: Impulse response functions (IRFs) of selected variables to a $1$ standard deviation (positive) TFP shock with $\chi=8.7360 \times 10^{-4}$.}
    \label{fig:IRFs_TFP_HighChi}
\end{sidewaysfigure}

\begin{sidewaystable}[p] \centering 
    \caption{Means - Capital adjustment costs}
    \begin{tabular}{lP{3.5cm}P{0.5cm}P{3.5cm}P{0.5cm}P{3cm}P{3cm}P{3cm}} 
    \hline \hline 
    Variable & BAU & & Unc. Opt. Policy & & \multicolumn{3}{c}{Con. Opt. Policy} \\ \cmidrule{2-2} \cmidrule{4-4} \cmidrule{6-8} & & & & & Rebate to S ($\xi=0$) & Uniform Red. ($\xi=\gamma$) & Red. to HtM ($\xi=1$) \\\cmidrule{6-8} 
    \multicolumn{4}{c}{} \\ 
    \hline 
    $\log(Y_{t})$ & $1.002$ & & $0.995$ & & $0.991$ & $0.995$ & $0.998$ \\ 
    $\log(I_{t})$ & $0.212$ & & $0.208$ & & $0.206$ & $0.208$ & $0.21$ \\ 
    $\log(f(\mu_{t})Y_{t})$ & $0$ & & $0.002$ & & $0.005$ & $0.002$ & $0.001$ \\ 
    $\log(C_{t})$ & $0.789$ & & $0.784$ & & $0.78$ & $0.784$ & $0.787$ \\ 
    \multicolumn{4}{c}{} \\ 
    $\log(C^{S}_{t})$ & $0.827$ & & $0.784$ & & $0.82$ & $0.82$ & $0.815$ \\ 
    $\log(C^{H}_{t})$ & $0.641$ & & $0.784$ & & $0.623$ & $0.643$ & $0.674$ \\ 
    \multicolumn{4}{c}{} \\ 
    $\log(V^{X}_{t})$ & $0.034$ & & $0.02$ & & $0.032$ & $0.031$ & $0.028$ \\ 
    $\log(\tau^{*}_{t})$ & $0$ & & $0.02$ & & $0.032$ & $0.02$ & $0.01$ \\ 
    $\log(E_{t})$ & $1.001$ & & $0.675$ & & $0.58$ & $0.675$ & $0.78$ \\ 
    $\log(\tau^{*}_{t}E_{t})$ & $0$ & & $0.014$ & & $0.019$ & $0.014$ & $0.008$ \\ 
    $\log(X_{t})$ & $476.65$ & & $321.45$ & & $276.07$ & $321.27$ & $371.52$ \\ 
    $\log(\mu_{t})$ & $0$ & & $0.322$ & & $0.416$ & $0.322$ & $0.218$ \\ 
    \multicolumn{4}{c}{} \\ 
    $\mathcal{W}_{t}$ & $-108.67$ & & $-70.003$ & & $-74.413$ & $-77.901$ & $-81.678$ \\ 
    $U_{t}^{S}$ & $-77.233$ & & $-70.003$ & & $-54.404$ & $-59.126$ & $-66.314$ \\ 
    $U_{t}^{H}$ & $-234.44$ & & $-70.003$ & & $-154.45$ & $-153$ & $-143.13$ \\ 
    \hline \hline 
    \end{tabular}
    \floatfoot{Notes: Averages for selected macroeconomic and environmental variables with $\epsilon=1.5$.}
    \label{tab:SimMeans_HighEps}
\end{sidewaystable}
    
\begin{sidewaystable}[p] \centering 
    \caption{Standard deviations ($\%$) - Capital adjustment costs}
    \begin{tabular}{lP{3.5cm}P{0.5cm}P{3.5cm}P{0.5cm}P{3cm}P{3cm}P{3cm}} 
    \hline \hline 
    Variable & BAU & & Unc. Opt. Policy & & \multicolumn{3}{c}{Con. Opt. Policy} \\ \cmidrule{2-2} \cmidrule{4-4} \cmidrule{6-8} 
    & & & & & Rebate to S ($\xi=0$) & Uniform Red. ($\xi=\gamma$) & Red. to HtM ($\xi=1$) \\ \cmidrule{6-8} 
    \multicolumn{4}{c}{} \\ \hline 
    $\log(Y_{t})$ & $3.13$ & & $3.08$ & & $3.04$ & $3.05$ & $3.04$ \\ 
    $\log(I_{t})$ & $5.43$ & & $5.23$ & & $5.00$ & $5.06$ & $5.03$ \\ 
    $\log(C_{t})$ & $2.51$ & & $2.49$ & & $2.47$ & $2.49$ & $2.50$ \\ 
    \multicolumn{4}{c}{} \\ 
    $\log(C^{S}_{t})$ & $2.39$ & & $2.49$ & & $2.39$ & $2.37$ & $2.29$ \\ 
    $\log(C^{H}_{t})$ & $3.13$ & & $2.49$ & & $2.97$ & $3.08$ & $3.51$ \\ 
    $\log(\lambda^{H}_{t})$ & $7.19$ & & $0.00$ & & $4.79$ & $6.07$ & $11.29$ \\ 
    \multicolumn{4}{c}{} \\ 
    $\log(\tau^{*}_{t})$ & $0.00$ & & $6.36$ & & $5.14$ & $7.09$ & $11.82$ \\ 
    $\log(E_{t})$ & $2.18$ & & $0.81$ & & $0.71$ & $0.62$ & $0.53$ \\ 
    $\log(V^{X}_{t})$ & $5.22$ & & $6.36$ & & $5.14$ & $5.14$ & $4.73$ \\
    \hline \hline
    \end{tabular}
    \floatfoot{Notes: Standard deviations for selected macroeconomic and environmental variables with $\epsilon=1.5$.}
    \label{tab:SimVola_HighEps}
\end{sidewaystable} 

\begin{sidewaysfigure}[p] \centering
    \caption{Responses to TFP shocks - Capital adjustment costs}
    \includegraphics[width=\textwidth,height=11cm,trim=2cm 0.4cm 2.5cm 0.5cm,clip]{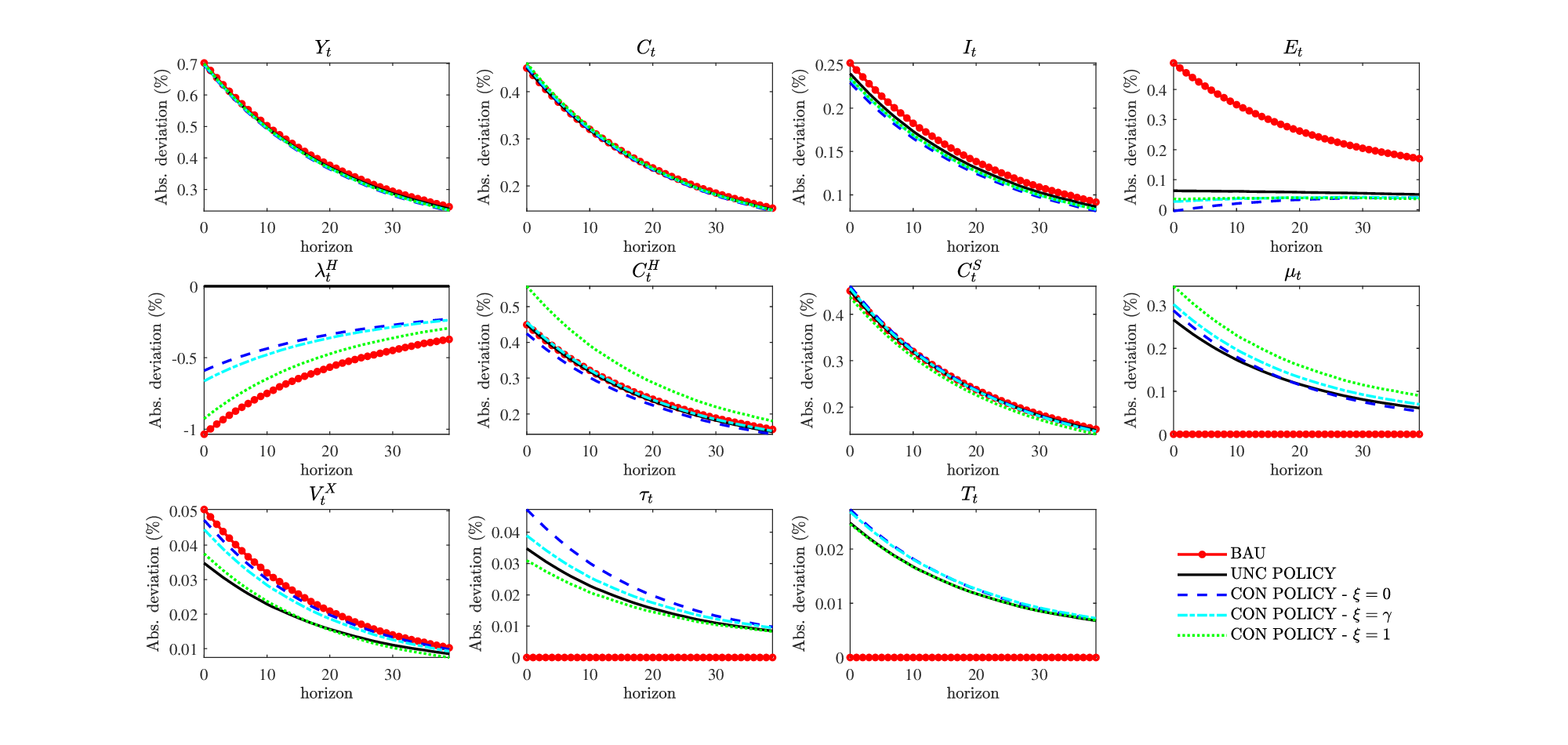}
    \vspace{-.7cm}
    \floatfoot{Notes: Impulse response functions (IRFs) of selected variables to a $1$ standard deviation (positive) TFP shock with $\epsilon=1.5$.}
    \label{fig:IRFs_TFP_HighEps}
\end{sidewaysfigure}




\end{appendices}

\end{document}